\documentclass[10pt,a4paper]{article}

\usepackage{natbib}
\usepackage{graphicx}
\usepackage{amssymb} 
\usepackage{amsmath}
\usepackage{subcaption}
\usepackage{float}

\begin{document}
	
	\title{The gradual evolution of \\ buyer--seller networks \\ and their role in aggregate fluctuations
	} 
	

	\author{Ryohei Hisano, Tsutomu Watanabe, Takayuki Mizuno, Takaaki Ohnishi \\ and Didier Sornette\thanks{Hisano: Social ICT Research Center, Graduate School of Information Science and Technology, The University of Tokyo, 7-3-1 Hongo, Bunkyo-ku, Tokyo 113-8654 Japan, em072010@yahoo.co.jp.  Watanabe: Graduate School of Economics, The University of Tokyo, 7-3-1 Hongo, Bunkyo-ku, Tokyo 113-8654 Japan, watanabe1284@gmail.com.  Mizuno: Information and Society Research Division, National Institute of Informatics, 2-1-2 Hitotsubashi, Chiyoda-ku, Tokyo 101-8430, mizuno@nii.ac.jp.  Ohnishi: Social ICT Research Center, Graduate School of Information Science and Technology, The University of Tokyo, 7-3-1 Hongo, Bunkyo-ku, Tokyo 113-8654 Japan, ohnishi.takaaki@i.u-tokyo.ac.jp.  Sornette: Department of Management, Technology and Economics, ETH Z\"{u}rich, Swiss Federal Institute of Technology, Scheuchzerstrasse 7, 8092 Z\"{u}rich, Switzerland, sornette@ethz.ch.  The authors are grateful to Hiroshi Iyetomi, Shoji Fujimoto and the seminar participants at RIETI for their helpful comments to the previous versions of the paper.  Hisano was supported by funding from the Research Fellowships of Japan Society for the Promotion of Science for Young Scientists.}}

	\date{\today}

	\maketitle
	

	\begin{abstract}
		Buyer--seller relationships among firms can be regarded as a longitudinal network in which the connectivity pattern evolves as each firm receives productivity shocks. Based on a data set describing the evolution of buyer--seller links among 55,608 firms over a decade and structural equation modeling, we find some evidence that interfirm networks evolve reflecting a firm's local decisions to mitigate adverse effects from neighbor firms through interfirm linkage, while enjoying positive effects from them.  As a result, link renewal tends to have a positive impact on the growth rates of firms.  We also investigate the role of networks in aggregate fluctuations.


	\end{abstract}
	
	The interfirm buyer--seller network is important from both the macroeconomic and the microeconomic perspectives.  From the macroeconomic perspective, this network represents a form of interconnectedness in an economy that allows firm-level idiosyncratic shocks to be propagated to other firms.  Previous studies has suggested that this propagation mechanism interferes with the averaging-out process of shocks, and possibly has an impact on macroeconomic variables such as aggregate fluctuations  (\cite{Acemoglu2013}, \cite{Acemoglu2012}, \cite{Carvalho2014}, \cite{Carvalho2007}, \cite{Shea2002}, \cite{Forester2011} and \cite{Malysheva2011}).  From the microeconomic perspective, a network at a particular point of time is a result of each firms link renewal decisions in order to avoid (or share) negative (or positive) shocks with its neighboring firms.  These two views of a network is related by the fact that both concerns propagation of shocks.  The former view stresses the fact that idiosyncratic shocks propagates through a static network while the latter provides a more dynamic view where firms have the choice of renewing its link structure in order to share or avoid shocks.  The question here is that it is not clear how the latter view affects the former view.  Does link renewal increase aggregate fluctuation due to firms forming new links that conveys positive shocks or does it decrease aggregate fluctuation due to firms severing links that conveys negative shocks or does it have a different effect?
	
	It is important to stress the fact that previous research, in macroeconomics as listed above, has implicitly assumed a static link structure where link renewal does not take place.  However, anecdotal evidence suggest that firms may renew their link structure in order to avoid negative shocks and share positive shocks with their neighboring firms.  For instance, in the financial crisis of 2008 many banks were reported to sever its links with bad performing firms while forming new links to better performing firms.  If these decisions took place broadly then shocks would not propagate as the previous  papers have suggested.

	To investigate the trade--off between the propagation of shocks and link renewal, we conduct an empirical analysis on the effect of link renewal on the overall growth rate of an economy. Our analysis is novel in the sense that we take the link renewal aspect of the network explicitly into account.  This is performed by employing a firm level data instead of sectoral level data.  Due to data availability, we use a firm-level dataset from Japan where we have both network data as well as log growth rate of each firms over a decade.  We hope that similar results holds for other countries as well.

	Using the unique dataset, we take structural equation modeling to estimate the effect of link renewal on the overall growth rate of a network.  Our model can be seen as a firm-level variant of the multi-sector model of \cite{Long1983}, which is canonical in the business-cycle literature.  After estimation of the structural parameters, wherein we discuss the results and identification issues, the effect of link renewal is estimated by performing a counterfactual analysis of the propagation of shocks.  Specifically, the analysis is performed by first estimating the individual shocks using the estimated structural model and then propagating the shocks back using networks from different years and comparing the consequences.  From this excercise our first result shows that the current network is often the best network configuration, which optimizes both the propagation of positive shocks and the avoidance of negative shocks compared with previous networks.  Furthermore, we show that for positive shocks, the future network is often better than the current network in the sense that it propagates positive shocks better than the current network.  This is explained by the asymmetry in cost between severing a link and link formation.  It is easier to sever an existing link when one's neighbor faces negative shocks than to form a new link, or a new path to distant targeted nodes, in the opposite case.  We then provide some evidence that link renewal has a positive effect of increasing the average growth rate of firms, thereby answering to the main question of the paper.  Finally, by comparing the average log growth rate for each year and the average individual shocks estimated from our model, we show that at least 37\% of the aggregate fluctuations can be explained by the network effect.

	The rest of the paper is organized as follows.  In Section 1, we summarize the basic notation used throughout the paper.  We also offer a brief description of the dataset used in the paper and provide a basic descriptive analysis.  Section 2 presents the structural model. Section 3 illustrates our inference procedure and presents the estimation results.  We also discuss identification issues.  In Section 4, we use the model to perform counterfactual analysis of the propagation of shocks and address the gradual evolution of the network.   Section 5 addresses the impact of the interfirm buyer--seller network on aggregate fluctuations.  Section 6 concludes.

	\section{Data and Notation}

	The network and financial data used in this paper are from the Teikoku Data  Bank\footnote{http://www.tdb.co.jp/index.html}.  These data are based on questionnaires completed by more than 100,000 firms in Japan for the accounting years 2003 to 2012.  We use a subset of this data where we have both network and financial information throughout the 10-year period (i.e., 55,608 firms).  In the questionnaires, firms are asked to name several (up to five) upstream and downstream firms with which they trade.  This scheme is akin to the fixed rank nomination scheme used in social network analysis (\cite{Hoff2013}).

	We define two types of adjacency matrix: downstream and upstream. We denote by $G$ the adjacency matrix describing the downstream network, where the downstream firms are listed in each row.  Thus, it is reported by firm $i$ that firm $j$ buys from firm $i$ if and only if $G_{ij} = 1$. $H$ is defined similarly for the upstream adjacency matrix. When necessary, we use subscripts to indicate time points, so the buyer network for accounting year 2012 is denoted by $G_{2012}$.  We could combine these two adjacency matrices and create matrices such that $H = G^{T}$ holds using interpolation of links.  However, because the data do not include the weight (i.e. transaction volume) spurious links might be formed using this interpolation.  To elaborate on this point, suppose that a stationery store sells a considerable number of pencils to firm A, which manufactures cars.  From the stationery store's point of view, firm A is a major buyer that determines its sales revenue.  However, from firm A's point of view, the stationery store is far less important than the upstream firm from which it purchases automobile parts for use in production.  Because in this paper we focus on links that have strong relationships, we focus on the raw form without performing any interpolation of relations.  It is worth noting that thus $G$ does not equal its transpose of $H$.

	Table 1 summarizes some basic descriptive statistics concerning the log growth rate of firms during the period 2003-2012.  It can be seen that the average log growth rate of firms fluctuates around 0, showing a moderate cycle.  As stated previously, because we are using a subset of the data, 55,608 firms were used to calculate the average log growth each year.  Table 2 summarizes the number of nonzero elements in the two adjacency matrices, as well as their evolution.  It can be seen that, except for 2008, the numbers of links formed and severed have shown a steady evolution.  It can also be seen that the overall number of links appears to be stable over time.

	\begin{table}[!htbp] \centering  
		\label{table:1}  
		\begin{tabular}{lccccc}
			\\[-1.8ex]\hline  
			\hline \\[-1.8ex]  
			Year & \multicolumn{1}{c}{Mean Log Growth} & \multicolumn{1}{c}{Standard Deviation}  \\  
			\hline \\[-1.8ex]  
			2003 & 0.008 & 0.182 \\ 
			2004 & 0.021 & 0.174 \\ 
			2005 & 0.022 & 0.161 \\ 
			2006 & 0.022 & 0.172 \\ 
			2007 & 0.017 & 0.175 \\ 
			2008 & 0.001 & 0.196 \\ 
			2009 & --0.076 & 0.220 \\ 
			2010 & --0.059 & 0.227 \\ 
			2011 & 0.009 & 0.206 \\ 
			2012 & 0.004 & 0.188 \\ 
			\hline \\[-1.8ex]  
		\end{tabular} 
		\caption{Average log growth rate of firms and standard deviation.}  
	\end{table}

	\begin{table}[!htbp] \centering  
		\label{table:2}  
		\begin{tabular}{@{\extracolsep{1pt}}lcccccc}  
			\\[-1.2ex]\hline  
			\hline \\[-1.2ex]  
			Year & \multicolumn{1}{c}{G} & \multicolumn{1}{c}{Form G} & \multicolumn{1}{c}{Sever G} & \multicolumn{1}{c}{H} & \multicolumn{1}{c}{Form H} & \multicolumn{1}{c}{Sever H} \\  
			\hline \\[-1.2ex]  
			2003 & 105,238 & - & - & 116,980 & - & - \\ 
			2004 & 106,230 & 16,789 & 17,781 & 118,534 & 16,270 & 17,824 \\ 
			2005 & 106,425 & 16,862 & 17,057 & 119,228 & 16,736 & 17,430 \\ 
			2006 & 106,758 & 16,056 & 16,389 & 119,571 & 16,318 & 16,661 \\ 
			2007 & 106,732 & 15,924 & 15,898 & 119,625 & 16,110 & 16,164 \\  
			2008 & 109,073 & 20,375 & 22,716 & 122,075 & 20,699 & 23,149 \\ 
			2009 & 109,881 & 16,680 & 17,488 & 122,898 & 16,519 & 17,342 \\ 
			2010 & 109,721 & 15,861 & 15,701 & 122,049 & 16,167 & 15,318 \\ 
			2011 & 109,546 & 15,404 & 15,229 & 122,021 & 14,675 & 14,647 \\ 
			2012 & 109,928 & 14,618 & 15,000 & 122,844 & 13,982 & 14,805 \\ 
			\hline \\[-1.2ex]  
		\end{tabular} 
		\caption{Number of nonzero elements in the two adjacency matrices and the number of new links (nonzero elements) formed and severed in the two matrices.}  
	\end{table}

\begin{figure}[!h]
		\begin{center} 
			\includegraphics*[width=.95\textwidth]{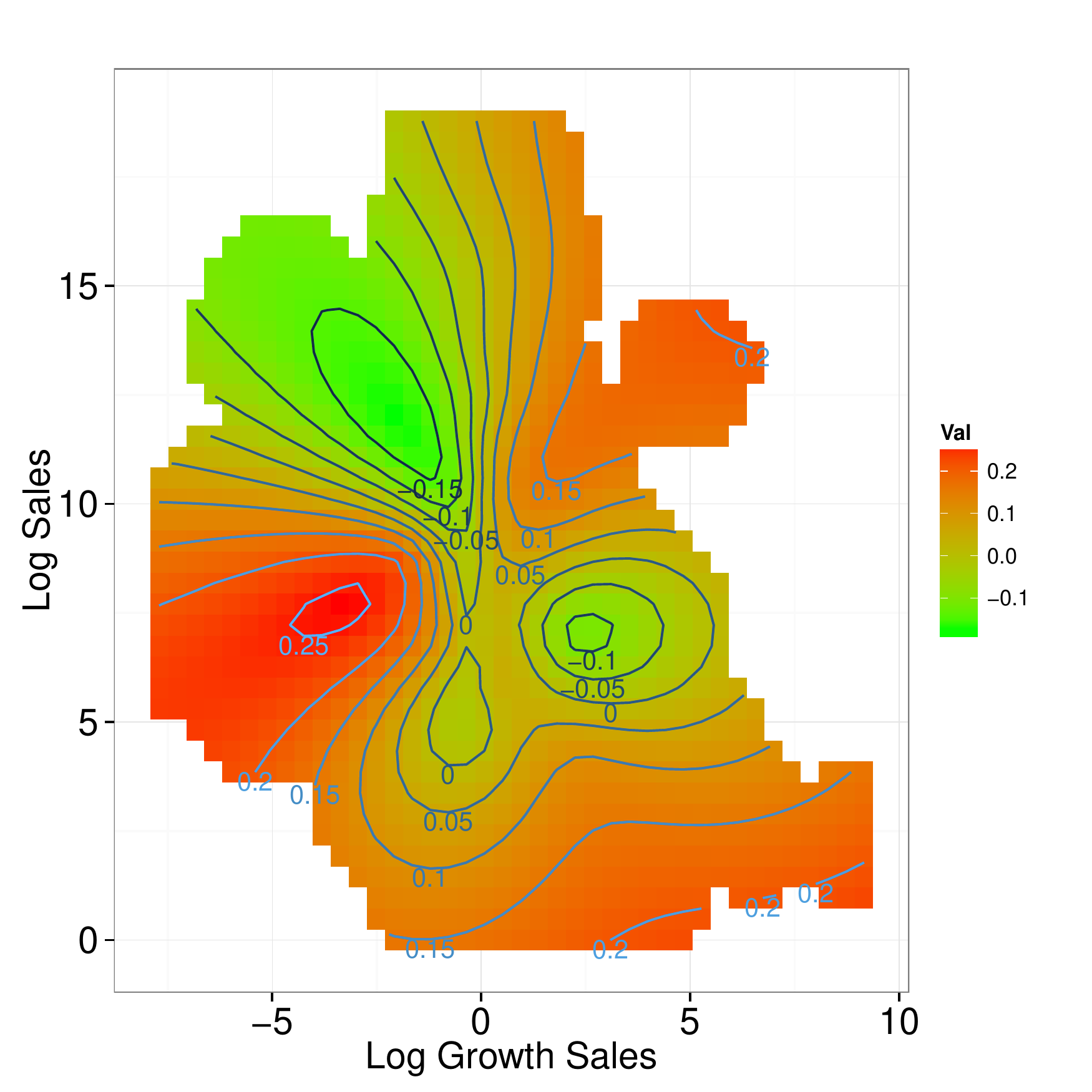} 
			\caption{(Color online) Contour plot showing the log growth rate of the following year to the current log growth rate and the current size for each firm.  The contour was estimated using two-dimensional splines.} 
			\label{fig:2} 
		\end{center} 
	\end{figure}

	In Figure~\ref{fig:2}, we present a contour plot showing the log growth rate of the following year (contour) to the current log growth rate (x-axis) and current size (y-axis) for each firm where the contour was estimated using two-dimensional splines. It can be seen that above 8.1 billion yen (i.e., $\exp(9)$), there is a clear persistent pattern whereby a positive growth rate tends to be repeated, and vice versa.  The irregular pattern seen around the middle left area could be explained by the behavior of subsidiary firms, which are affected by decisions made by their parent company (e.g., participating in an absorption-type merger, corporate group restructuring).  However, even ignoring this part of the data, it can be seen that overall, there seems to be a persistent pattern in the log growth rate of firms.

\begin{figure}[!h]
		\begin{center} 
			\includegraphics*[width=.95\textwidth]{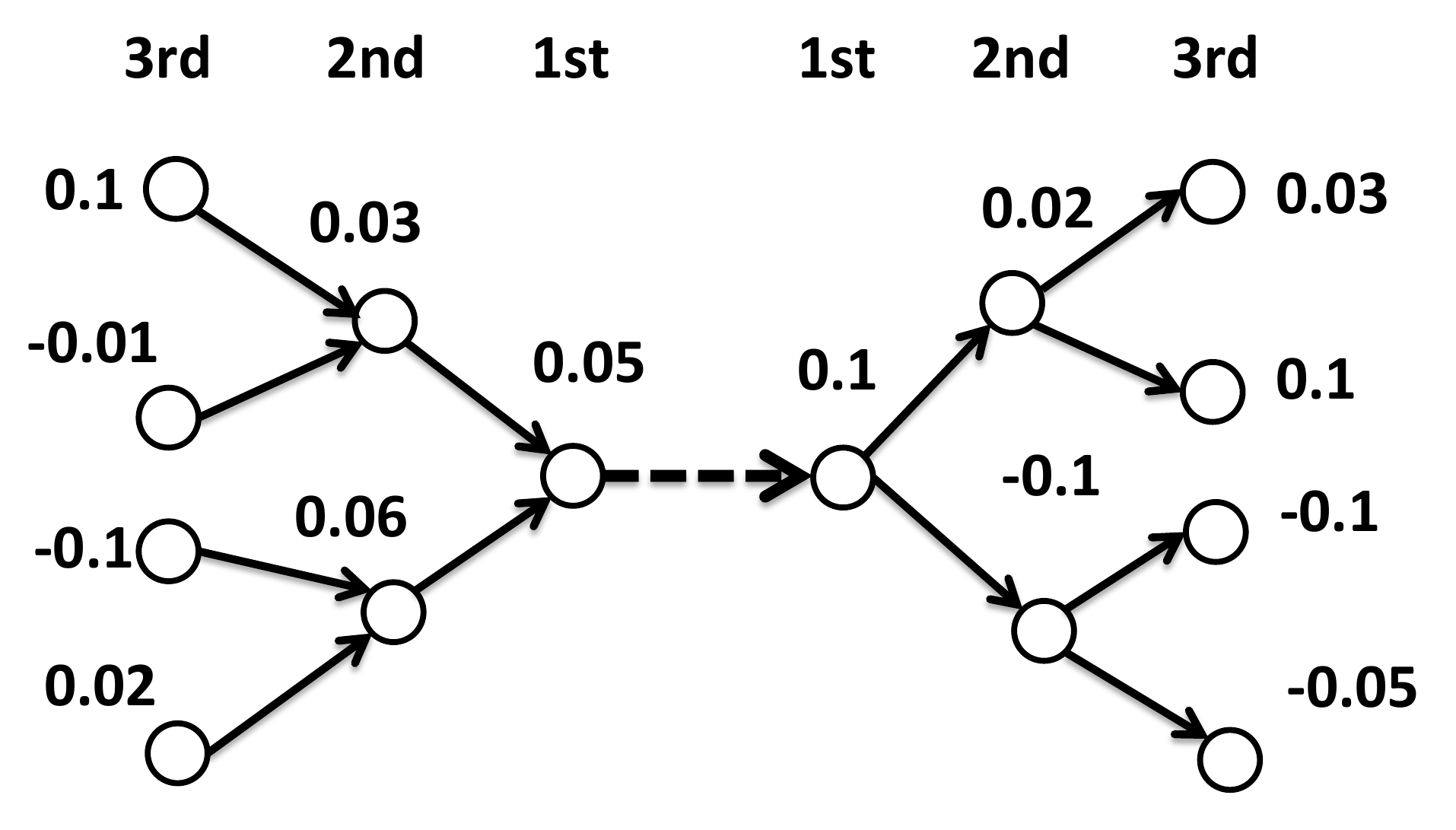} 
			\caption{Schematic diagram showing first-order, second-order, and third-order nodes of formed and severed links.  Dashed lines indicate newly formed or severed links.  The numbers represent the log growth rate of each firm.} 
			\label{fig:3} 
		\end{center} 
	\end{figure}

	In Table 3, we show the proportions of positive and negative log growth rates of firms around newly formed and severed links.  First-order, second-order, and third-order nodes are defined by the steps needed to reach the node from the newly formed or severed link.  For the sake of clarity, a schematic diagram showing the first-order, second-order, and third-order nodes is provided in figure~\ref{fig:3}.  Bold font in Table 3 indicates the cases where (i) the proportion of positive log growth rate of nodes is higher for newly formed links than severed links or (ii) the proportion of negative log growth rate of nodes is higher for severed links than newly formed links in a given year.  It can be seen that for all years, the network tends to form links between nodes experiencing a positive log growth rate (and vice versa).  This provides our first insight into the connection between the log growth rate of firms and the link renewal process of the network.

	\begin{table}[!tbp] 
		\begin{center} 
			\begin{tabular}{lccccccc} 
				\hline\hline
				\multicolumn{1}{c}{Year}
				&\multicolumn{1}{c}{Type}
				&\multicolumn{1}{c}{1st Pos}
				&\multicolumn{1}{c}{1st Neg}
				&\multicolumn{1}{c}{2nd Pos}
				&\multicolumn{1}{c}{2nd Neg}
				&\multicolumn{1}{c}{3rd Pos}
				&\multicolumn{1}{c}{3rd Neg}\tabularnewline 
				\hline 
				
				$2004$&$Sever$&$\bf{0.558}$&$\bf{0.423}$&$\bf{0.576}$&$\bf{0.421}$&$\bf{0.565}$&$\bf{0.432}$\tabularnewline 
				$2004$&$Form$ &$\bf{0.598}$&$\bf{0.388}$&$\bf{0.593}$&$\bf{0.404}$&$\bf{0.579}$&$\bf{0.419}$\tabularnewline	
				
				$2005$&$Sever$&$\bf{0.643}$&$\bf{0.333}$&$\bf{0.743}$&$\bf{0.25}$&$\bf{0.76}$&$\bf{0.233}$\tabularnewline 
				$2005$&$Form$ &$\bf{0.668}$&$\bf{0.315}$&$\bf{0.757}$&$\bf{0.235}$&$\bf{0.762}$&$\bf{0.231}$\tabularnewline	
				
				$2006$&$Sever$&$\bf{0.649}$&$\bf{0.324}$&$0.751$&$0.244$&$\bf{0.755}$&$\bf{0.241}$\tabularnewline 
				$2006$&$Form$ &$\bf{0.666}$&$\bf{0.316}$&$0.75$&$0.245$&$\bf{0.757}$&$\bf{0.238}$\tabularnewline	
				
				$2007$&$Sever$&$\bf{0.651}$&$\bf{0.319}$&$\bf{0.741}$&$\bf{0.252}$&$0.754$&$0.241$\tabularnewline 
				$2007$&$Form$ &$\bf{0.664}$&$\bf{0.317}$&$\bf{0.753}$&$\bf{0.242}$&$0.753$&$0.241$\tabularnewline		
				
				$2008$&$Sever$&$\bf{0.577}$&$\bf{0.396}$&$\bf{0.736}$&$\bf{0.258}$&$\bf{0.766}$&$\bf{0.229}$\tabularnewline 
				$2008$&$Form$ &$\bf{0.59}$&$\bf{0.393}$&$\bf{0.745}$&$\bf{0.249}$&$\bf{0.771}$&$\bf{0.225}$\tabularnewline
				
				$2009$&$Sever$&$\bf{0.266}$&$\bf{0.713}$&$0.293$&$0.704$&$0.294$&$0.704$\tabularnewline 
				$2009$&$Form$ &$\bf{0.284}$&$\bf{0.702}$&$0.293$&$0.704$&$0.286$&$0.712$\tabularnewline
				
				$2010$&$Sever$&$\bf{0.267}$&$\bf{0.705}$&$0.179$&$0.812$&$\bf{0.153}$&$\bf{0.836}$\tabularnewline 
				$2010$&$Form$ &$\bf{0.297}$&$\bf{0.687}$&$0.178$&$0.813$&$\bf{0.159}$&$\bf{0.832}$\tabularnewline
				
				$2011$&$Sever$&$\bf{0.567}$&$\bf{0.407}$&$\bf{0.683}$&$\bf{0.312}$&$\bf{0.716}$&$\bf{0.28}$\tabularnewline 
				$2011$&$Form$ &$\bf{0.597}$&$\bf{0.387}$&$\bf{0.703}$&$\bf{0.292}$&$\bf{0.731}$&$\bf{0.265}$\tabularnewline
				
				$2012$&$Sever$&$\bf{0.549}$&$\bf{0.416}$&$\bf{0.563}$&$\bf{0.427}$&$0.568$&$0.423$\tabularnewline 
				$2012$&$Form$&$\bf{0.585}$&$\bf{0.393}$&$\bf{0.566}$&$\bf{0.424}$&$0.558$&$0.433$\tabularnewline

				\hline 
			\end{tabular} 
			\label{table::3} 
			\caption{Proportions of positive and negative log growth rates of firms around a newly formed or severed link.  First-order, second-order, and third-order nodes are defined by the length of the newly formed or severed link.} 
		\end{center} 
	\end{table}

	\section{Model} 
	
	The model that we use in this paper is 
	\begin{eqnarray} 
	\left( I-\beta_{G}G_{t}-\beta_{H}H_{t} \right)y_{t}=\left( \beta  
	_{LG}G_{t-1}+\beta_{LH}H_{t-1} \right)y_{t-1}+\gamma y_{t-1}+\epsilon_{t}, 
	\end{eqnarray} 
	where $y_{t}$ denotes the growth rate of sales of each firm\footnote{Which is defined by the difference of logarithm of sales between two consecutive years.} and $\epsilon_{t}$ denotes the normal firm-specific idiosyncratic shock characterized by $\mu$ and $\sigma$.  There are seven unknown parameters in total. 
	
	Our model can be seen as a firm-level variant of the multi-sector model of \cite{Long1983}, which is canonical in the business-cycle literature.  In \cite{Long1983}, each sector (i.e., firm) is explicitly assumed to use materials produced by other sectors (i.e., firms), and these sectoral linkages represent interconnectedness in the economy, propagating idiosyncratic sector-specific shocks to other sectors.  The difference between \cite{Long1983}'s sectoral-level and firm-level linkages lies in the link renewal process among firms.  In a sectoral-level setting, if the total demand for goods from other sectors is kept the same, then the strength of the links with other sectors does not change.  However, even in this case, the interfirm network structure might differ due to link renewal behaviors at the firm level.  Our main goal in this is paper is to take this link renewal behavior explicitly into account.

	The general consensus in macroeconomics has been that sector-specific shocks should average out over the entire economy based on Lucas's ``diversification argument" (\cite{Lucas1977}).  However, this view has recently been challenged from the network perspective by several authors (\cite{Shea2002}, \cite{Acemoglu2012}, \cite{Acemoglu2013} and \cite{Carvalho2007}) suggesting that in the presence of certain sectoral network structures, this argument may not apply.  In particular, \cite{Acemoglu2012} has shown that the rate of decay in aggregate fluctuations depends on the network structure governing interdependency among sectors.  Our model is closely related to \cite{Acemoglu2012}, but much closer to \cite{Shea2002}  in that we model effects from both upstream and downstream linkages.  Our work is also related in spirit to \cite{Forester2011} and \cite{Malysheva2011} in providing a systematic econometric analysis of the propagation of shocks and the relationship to aggregate fluctuations.  The difference is that while \cite{Forester2011} and \cite{Malysheva2011} focus on sectorial linkages, we focus more on micro connections in interfirm networks.

	\section{Estimation}

	\subsection{Parameter estimation} 
	Inference of parameters is most easily performed using Bayesian inference 
	(\cite{Westveld2011} and \cite{Goldsmith-Pinkham_Imbens2013}).  In our case, this 
	is also due to the heavy computation involved in handling large amounts of network data.  Using equation (1) and placing conjugate normal priors on $\beta_{G}, \beta_{H}, \beta_{LG}, 
	\beta_{LH}, \gamma,$ and $\mu_{0}$, and a scaled inverse gamma prior on $\sigma_{0}$, 
	$y_{t}$ obeys a multivariate normal distribution with  
	\begin{eqnarray} 
	\mu =\left( I-\beta_{G}G-\beta_{H}H \right)^{-1}(\mu_{0}+\left( \beta 
	_{LG}G_{t-1}+\beta_{LH}H_{t-1}+\gamma I \right)y_{t-1}), \\ 
	\quad\Sigma =\left( I-\beta_{G}G-\beta_{H}H \right)^{-1}\left( I-\beta 
	_{G}G^{'}-\beta_{H}H^{'} \right)^{-1}\sigma_{0}^{2}. 
	\end{eqnarray}

	To perform maximum likelihood in this setting, it is necessary to calculate the determinant $|\Sigma|$, where $\Sigma$ has size $55,608^{2}$ even when focusing our attention on just one year.  The time complexity of calculating this determinant is cubic, making it impractical to evaluate when optimizing the likelihood\footnote[5]{It took about 5--8 hours to calculate this term on a modern desktop computer using the fully optimized software \cite{ARPACK++}}.  The other term that involves heavy computation is the inverse matrix. We approximated the inverse matrix using the first 30 terms of the Neumann series (or power series) as in \cite{Bramoulle2009}.

	The unknown parameters in our model are $\beta_{G}, \beta_{H}, \beta_{LG}, \beta_{LH}, \gamma, \mu_{0},$ and $\sigma_{0}$.  Bayesian inference was performed using Gibbs sampling of 10 years of data, which converged quite rapidly.  A Markov chain of 10,500 iterations was generated, the first 500 of which were dropped as burn-in steps.  Thinning was performed every 10 steps, resulting in 1,000 samples, which we used to approximate the joint posterior.  
	

	Table 4 reports the posterior mean of the parameters along with 99\% posterior confidence intervals.  In general, all the parameters related to network effects are significantly different from 0, suggesting that the network effect is present as both a lag and a contemporaneous effect.  The parameter $\gamma$ being significantly positive implies that there is persistency in firms log growth rate as was expected from figure 2.  The parameter $\mu_{0}$ being slightly negative corresponds to the fact the overall Japan was shrinking during the period of analysis.

	\begin{table}[!tbp] 
		\begin{center} 
			\begin{tabular}{lrrr} 
				\hline\hline 
				\multicolumn{1}{c}{Parameter}&\multicolumn{1}{c}{Mean}&\multicolumn{1}{c}{Lower}& \multicolumn{1}{c}{Upper}\tabularnewline 
				\hline 
				$\beta_{G}$&$ 0.06217$&$ 0.06216$&$ 0.06218$\tabularnewline 
				$\beta_{H}$&$ 0.05179$&$ 0.05178$&$ 0.0518$\tabularnewline 
				$\beta_{LG}$&$ 0.001$&$ 0.001$&$ 0.001$\tabularnewline 
				$\beta_{LH}$&$ 0.0088$&$ 0.0088$&$ 0.0088$\tabularnewline 
				$\gamma$&$0.0188$&$0.0187$&$0.0188$\tabularnewline 
				$\mu_{0}$&$-0.0035$&$-0.0035$&$-0.0035$\tabularnewline 
				$\sigma_{0}$&$ 0.5365$&$ 0.5351$&$ 0.5379$\tabularnewline 
				\hline 
			\end{tabular} 
			\label{table:4} 
			\caption{Parameter estimates.  Posterior mean and 99\% posterior confidence intervals are reported.} 
		\end{center} 
	\end{table}

	\subsection{Identification issues resulting from measurement errors} 
	
	Although the use of the log growth rate in analyzing network effects is due to stationarity concerns log differencing makes each variables noisier.  Moreover sloppy reporting by small and medium-sized firms also contaminates the variable with additional measurement errors.  Estimation of true regression parameters when all measurements have additional noise was studied by Frisch in the 1930s under the rubric of statistical confluence analysis (\cite{Frisch1934} and \cite{Hendry_Morgan1989}).  Similar to its modern descendant, partial identification (\cite{Manski2009} and \cite{Tamer2010}), our results show that estimation of the structural parameters ignoring measurement error provides lower bounds on estimates of the true structural parameters.

	While this argument may seem trivial at first, it is important when we estimate the effects of the interfirm buyer--seller network on aggregate fluctuations in Section 6.  As noted in the Introduction, since our interest is in aggregate fluctuation we are interested not in each firm's log growth rate, but in the average log growth rate of all firms in an economy at a particular year.  Additional zero mean measurement errors for each firm disappear when we take the average of these growth rates, and thus have no impact on the overall dynamics of the average log growth rate.  However, we are trying to estimate these underlying parameters from log growth rates including additional measurement errors.  In this case, our estimated parameters (e.g., the parameter estimates reported in Table 4) would be different from the true structural parameters responsible for generating the aggregate fluctuations in the average log growth rate of firms.

	Taking measurement errors into account, our observed log growth rate of firms is generated from 
	\begin{eqnarray} 
	\left( I-\beta_{G}G_{t}-\beta_{H}H_{t} \right)z_{t}=\left( \beta  
	_{LG}G_{t-1}+\beta_{LH}H_{t-1} \right)z_{t-1}+\gamma z_{t-1}+\epsilon_{t}, 
	\end{eqnarray} 
	\begin{eqnarray} 
	y_{t} = z_{t} + \eta_{t}, 
	\end{eqnarray} 
	where the first equation models the network effect as in equation (1) and the second one models additional measurement errors.  Assuming that $\eta$ has mean 0 and a finite first moment, the law of large numbers guarantees that this additional measurement error cancels out in the aggregate.

	Assuming that both $\epsilon_{t}$ and $\eta_{t}$ are normally distributed random variables, it is obvious that there is a simple relationship between the parameter estimates ignoring this additional structure and the true parameters.  The relationship is  
	\begin{eqnarray} 
	\theta_{apparent}=r*\theta_{true}, 
	\end{eqnarray} 
	where $r$ is defined as  
	\begin{eqnarray} 
	r:=\frac{var(\epsilon_{t})}{var(\epsilon_{t})+var(\eta_{t})}. 
	\end{eqnarray} 
	Hence, our parameter estimates ignoring measurement errors, as in Table 4, give a scaled estimate of the true parameters.

	This effect is confirmed by the following experiments.  We first generate the underlying true log growth rates of firms using the actual network data with $\beta_{G}=0.06$, $\beta_{H}=0.06$, 
	$\beta_{LG}=0.04$, $\beta_{LH}=0.04$, $\gamma=-0.3$, $\mu=0$, and $\sigma=0.3$.  Then, for each firm, we add additional noise $\eta \sim  normal(0,0.15)$.  Table 8 reports the posterior means of parameter estimates with and without this additional noise.  We see that the parameters are scaled as predicted by equation (7).  
	
	In summary, the analysis performed in this section have clarified that the estimated structural parameters only provide a lower bound on the true parameters.  This was a result of identification issues concerning measurement errors.  Hence the message here is that our evaluation of propagation of shocks, performed in the next sections using the estimated parameters, could only be seen as a lower bound concerning the true level of propagation in an economy.

	\begin{table}[!tbp] 
		\begin{center} 
			\begin{tabular}{lrrrrr} 
				\hline\hline 
				\multicolumn{1}{c}{Type of noise}&\multicolumn{1}{c}{$\beta_{G}$}&\multicolumn{1}{c}{$\beta_{H}$}&\multicolumn{1}{c}{$\beta_{LG}$}&\multicolumn{1}{c}{$\beta_{LG}$}&\multicolumn{1}{c}{$\gamma$}\tabularnewline 
				\hline 
				No error&$0.065$&$0.068$&$0.04$&$0.042$&$-0.29$\tabularnewline 
				Normal error &$0.056$&$0.055$&$0.031$&$0.030$&$-0.237$\tabularnewline 
				\hline 
			\end{tabular} 
			\caption{Parameter estimates with measurement errors.  The true parameters are 
				reported in the text.} 
		\end{center} 
	\end{table}

	\section{Counterfactual Analysis of Propagation of Shocks} 
	
	To assess the nature of the evolving network, we perform counterfactual analysis of the propagation of shocks.  We do this by the following procedure.  Using a structural model describing the interfirm buyer--seller network, we estimate the structural firm-specific shocks for year $t$ as
	
	\begin{eqnarray} 
	e_{t} := ( I-\beta_{G}G_{t}-\beta_{H}H_{t} )y_{t} 
	\end{eqnarray}

	\noindent
	where $\beta_{G}$ and $\beta_{H}$ are parameters, $e_{t}$ and $y_{t}$ are vectors, and the rest matrices.  Using these estimates for all firms, we compute a firm's growth in a counterfactual world, assuming that the structure of the network is that of year $t'$ instead of year $t$ by
	\begin{eqnarray} 
	y_{t'|t} := ( I-\beta_{G}G_{t'}-\beta_{H}H_{t'} )^{-1}e_{t}. 
	\end{eqnarray} 
	
	Note that $y_{t|t}$ (i.e., propagating shocks using the network from the same year as the log growth rate) is the same as $y_{t}$.  Comparing $y_{t'|t}$ for different years enables us to ascertain what the log growth rate of firms might have been if the network structure was that of year $t'$.  Moreover, motivated by Table 3, we perform this analysis of evolving networks by separating the estimated $e_{t}$s into positive shocks  (i.e., $e_{t}^{pos}$) and negative shocks (i.e., $e_{t}^{neg}$) where we set all the values that are not positive in the former case or negative in the latter case to 0.  We propagate each of these shocks in the network.  Thus, $y_{t'|t}$ is now replaced by

	\begin{eqnarray} 
	y_{t'|t}^{pos} := ( I-\beta_{G}G_{t'}-\beta_{H}H_{t'} )^{-1}e_{t}^{pos}
	\end{eqnarray} 
	
	\noindent for positive shocks and
	
	\begin{eqnarray} 
	y_{t'|t}^{neg} := ( I-\beta_{G}G_{t'}-\beta_{H}H_{t'} )^{-1}e_{t}^{neg}
	\end{eqnarray} 
	
	\noindent for negative shocks.  We assume that the structural parameters are fixed and set them as $\beta_{G}=0.06$ and $\beta_{H}=0.05$. 
	
\begin{figure}[!h]
		\begin{minipage}[b]{0.243\linewidth}
			\centering
			\includegraphics[keepaspectratio, scale=0.2]
			{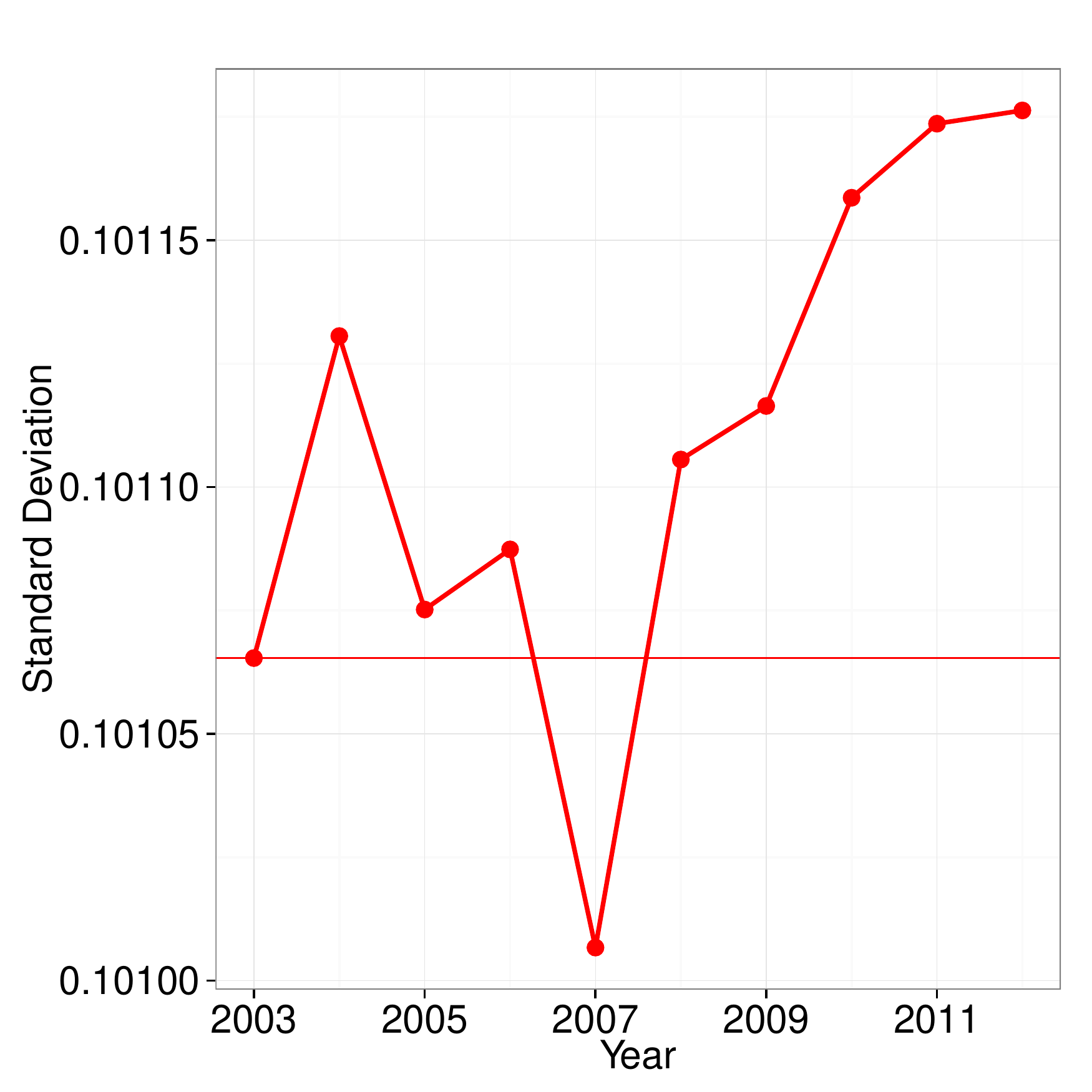}
			\subcaption{2003}
			\label{fig_neg_2003}
		\end{minipage}
		\begin{minipage}[b]{0.243\linewidth}
			\centering
			\includegraphics[keepaspectratio, scale=0.2]
			{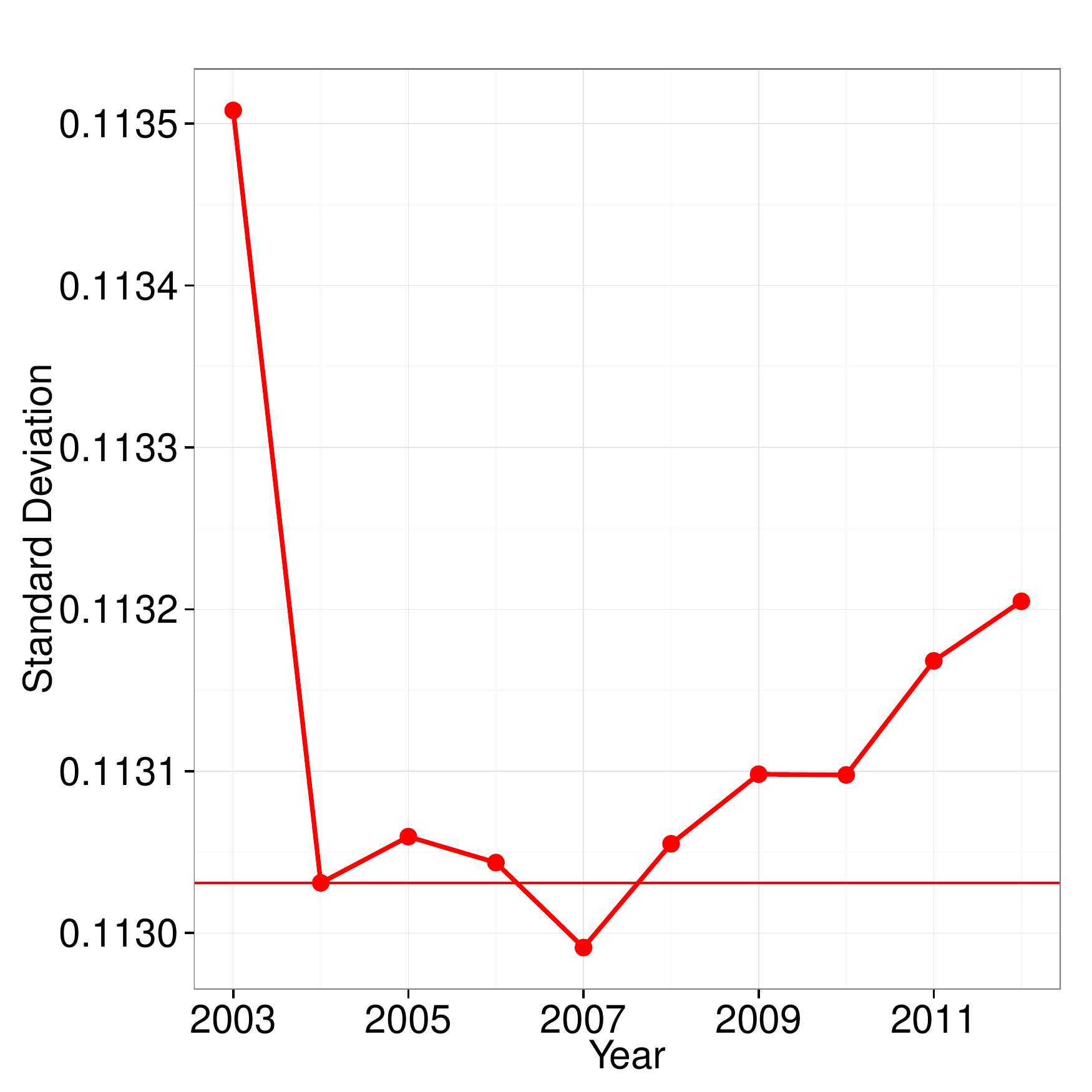}
			\subcaption{2004}
			\label{fig_neg_2004}
		\end{minipage}
		\begin{minipage}[b]{0.243\linewidth}
			\centering
			\includegraphics[keepaspectratio, scale=0.2]
			{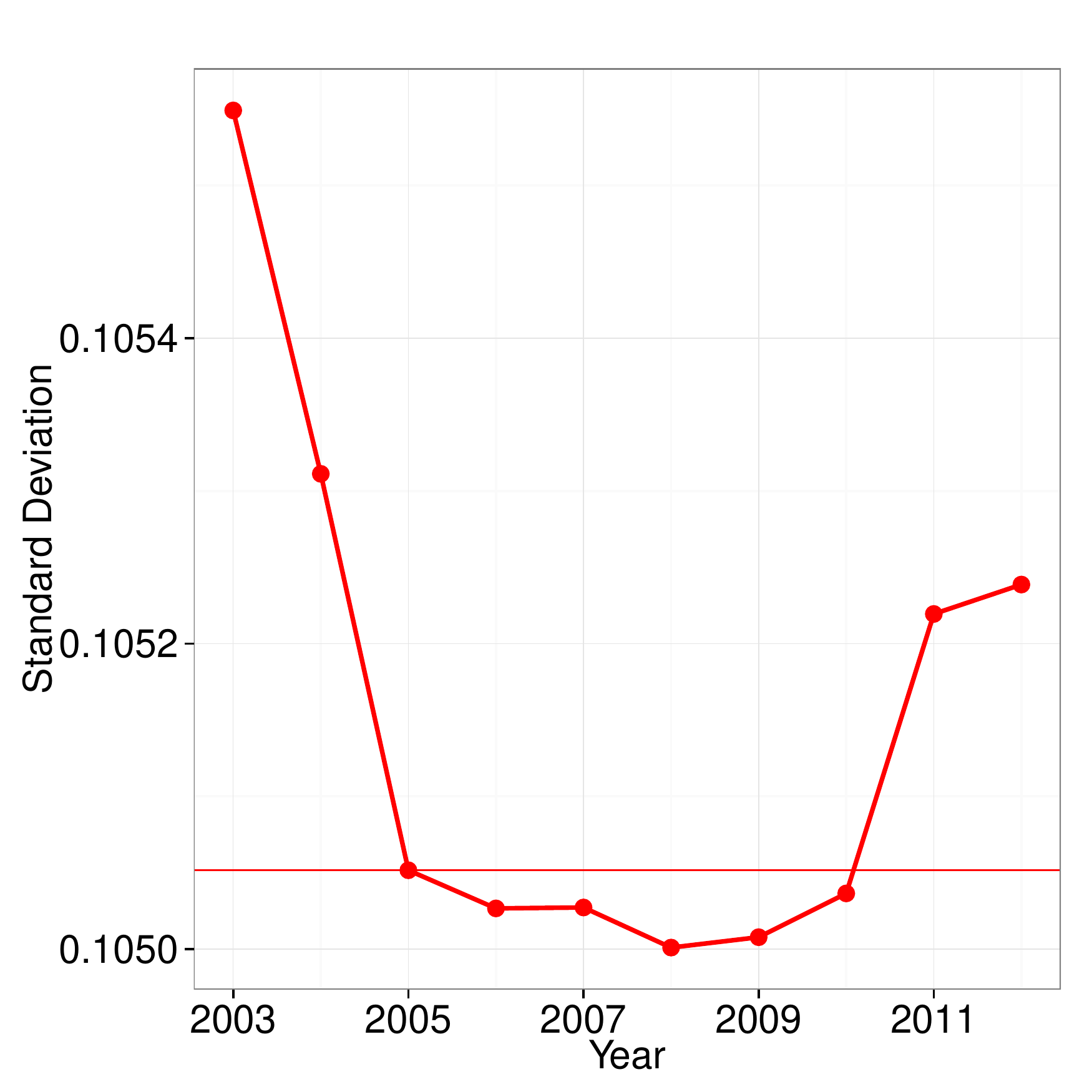}
			\subcaption{2005}
			\label{fig_neg_2005}
		\end{minipage}
		\begin{minipage}[b]{0.243\linewidth}
			\centering
			\includegraphics[keepaspectratio, scale=0.2]
			{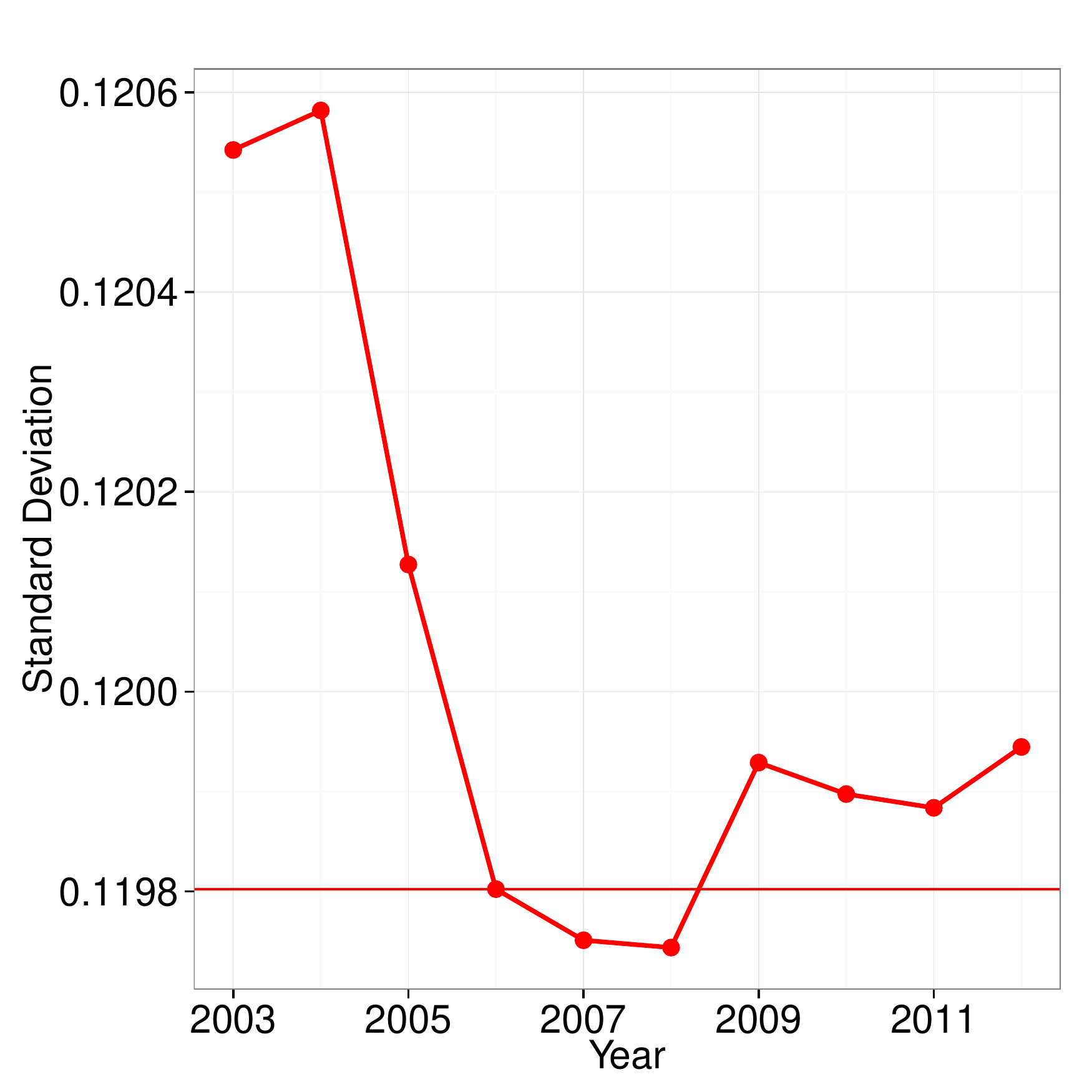}
			\subcaption{2006}
			\label{fig_neg_2006}
		\end{minipage}\\
		\begin{minipage}[b]{0.243\linewidth}
			\centering
			\includegraphics[keepaspectratio, scale=0.2]
			{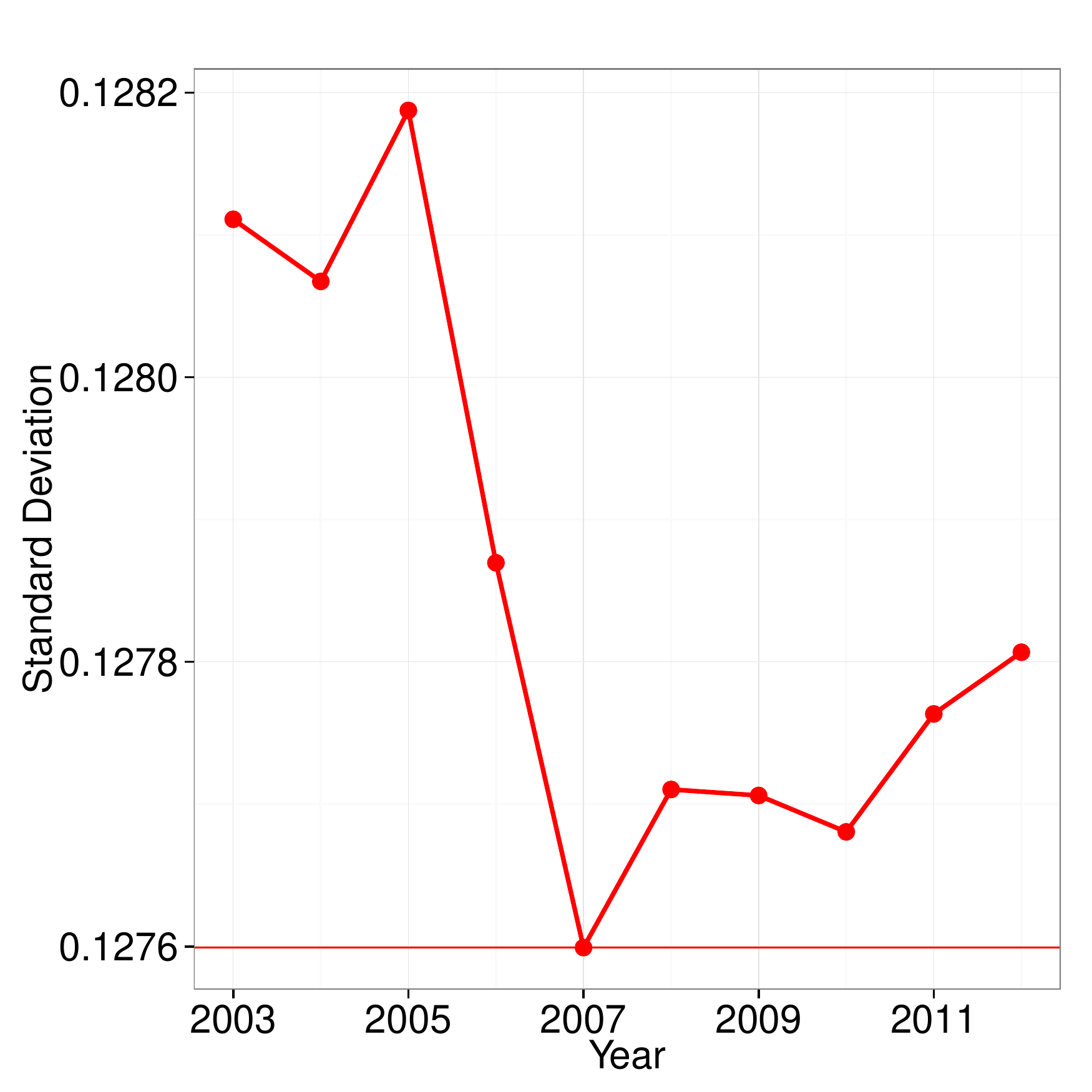}
			\subcaption{2007}
			\label{fig_neg_2007}
		\end{minipage}
		\begin{minipage}[b]{0.243\linewidth}
			\centering
			\includegraphics[keepaspectratio, scale=0.2]
			{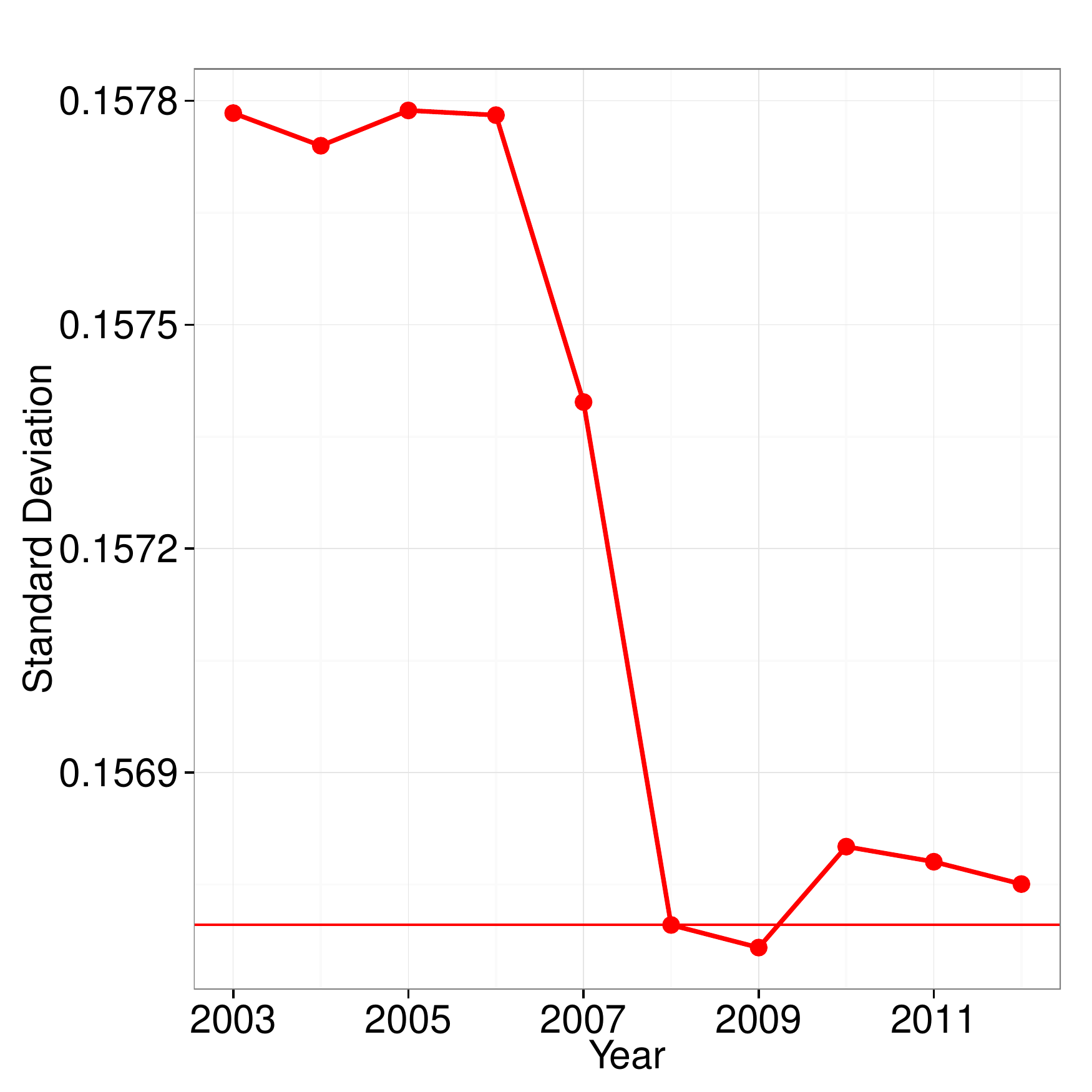}
			\subcaption{2008}
			\label{fig_neg_2008}
		\end{minipage}
		\begin{minipage}[b]{0.243\linewidth}
			\centering
			\includegraphics[keepaspectratio, scale=0.2]
			{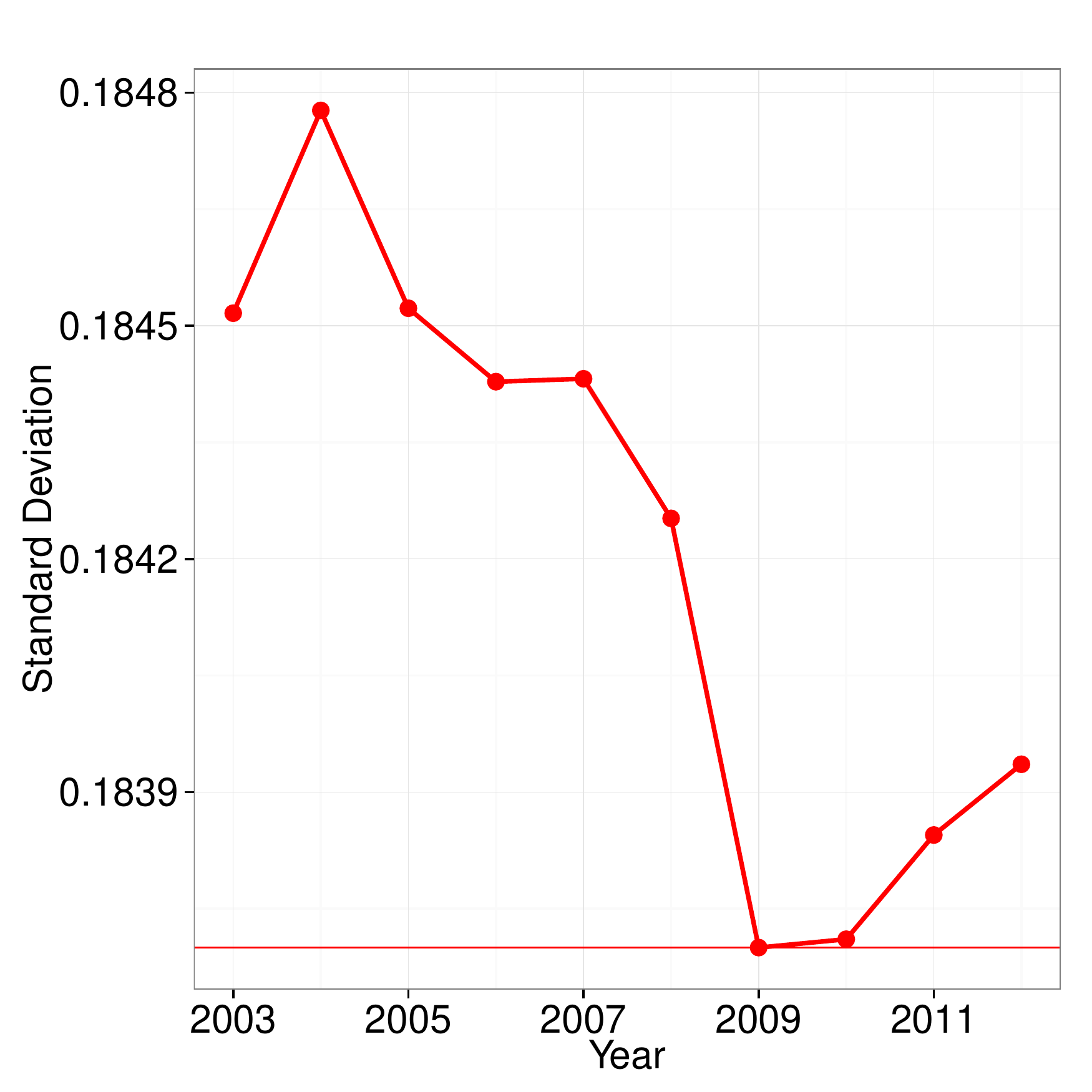}
			\subcaption{2009}
			\label{fig_neg_2009}
		\end{minipage}
		\begin{minipage}[b]{0.243\linewidth}
			\centering
			\includegraphics[keepaspectratio, scale=0.2]
			{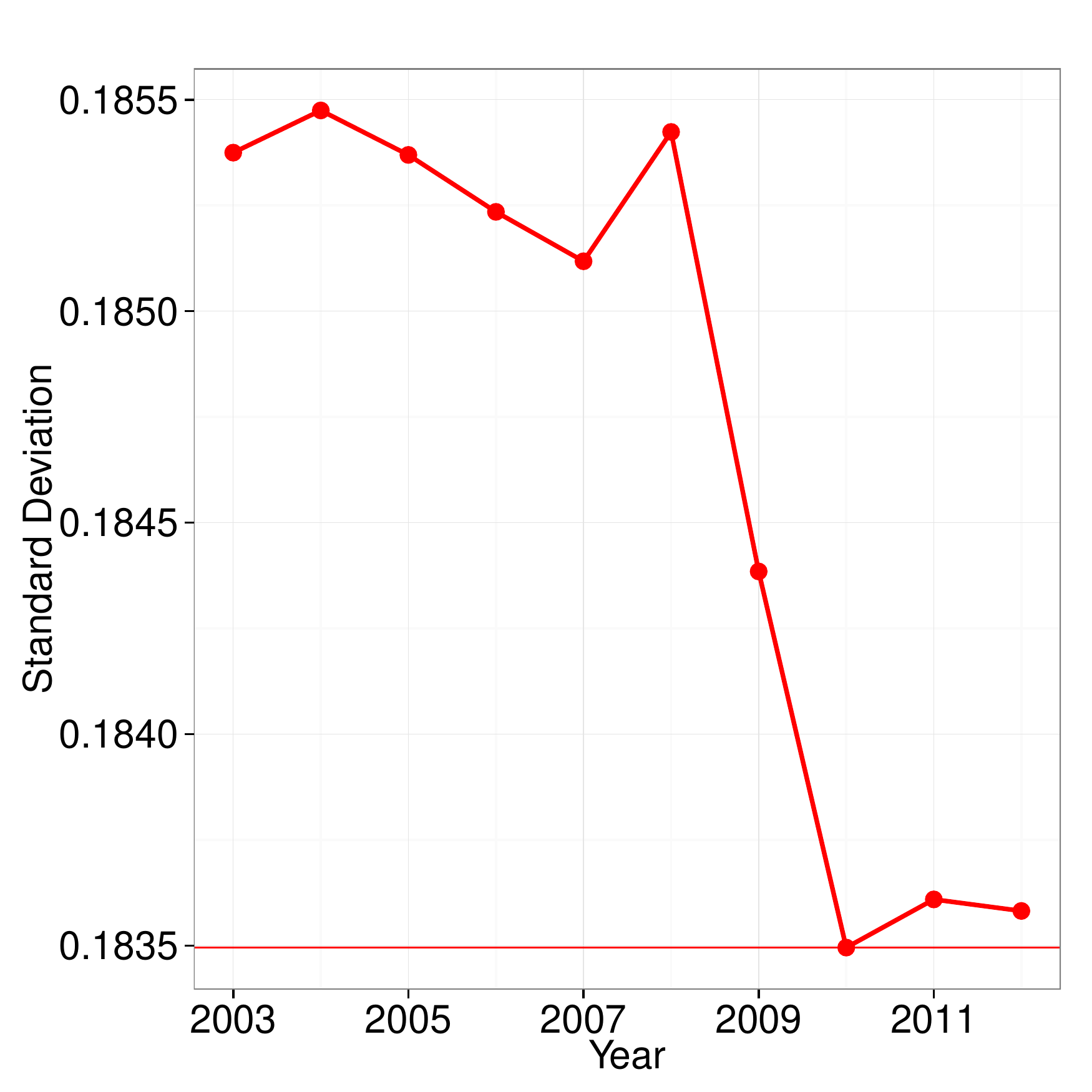}
			\subcaption{2010}
			\label{fig_neg_2010}
		\end{minipage}\\
		\begin{minipage}[b]{0.243\linewidth}
			\centering
			\includegraphics[keepaspectratio, scale=0.2]
			{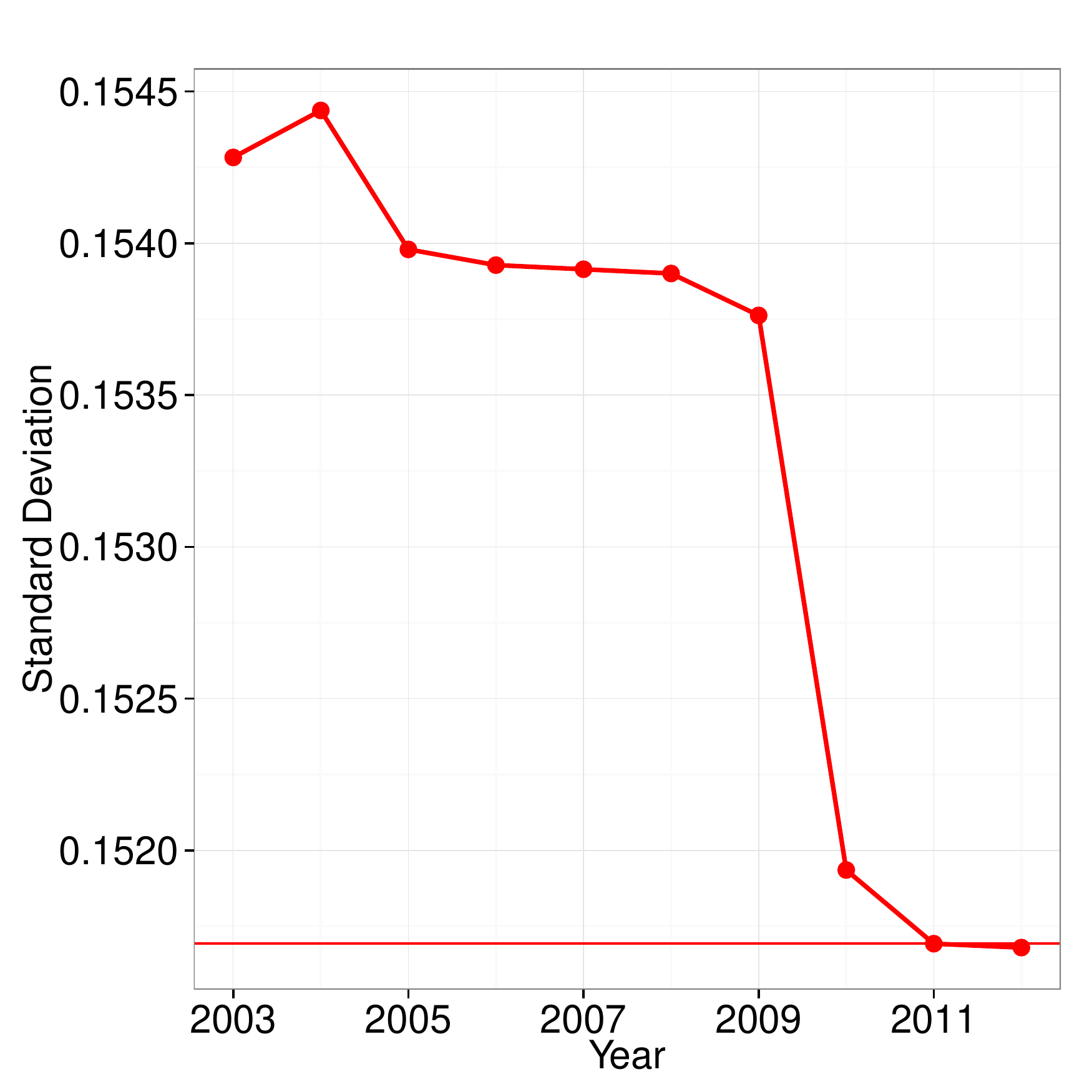}
			\subcaption{2011}
			\label{fig_neg_2011}
		\end{minipage}
		\begin{minipage}[b]{0.243\linewidth}
			\centering
			\includegraphics[keepaspectratio, scale=0.2]
			{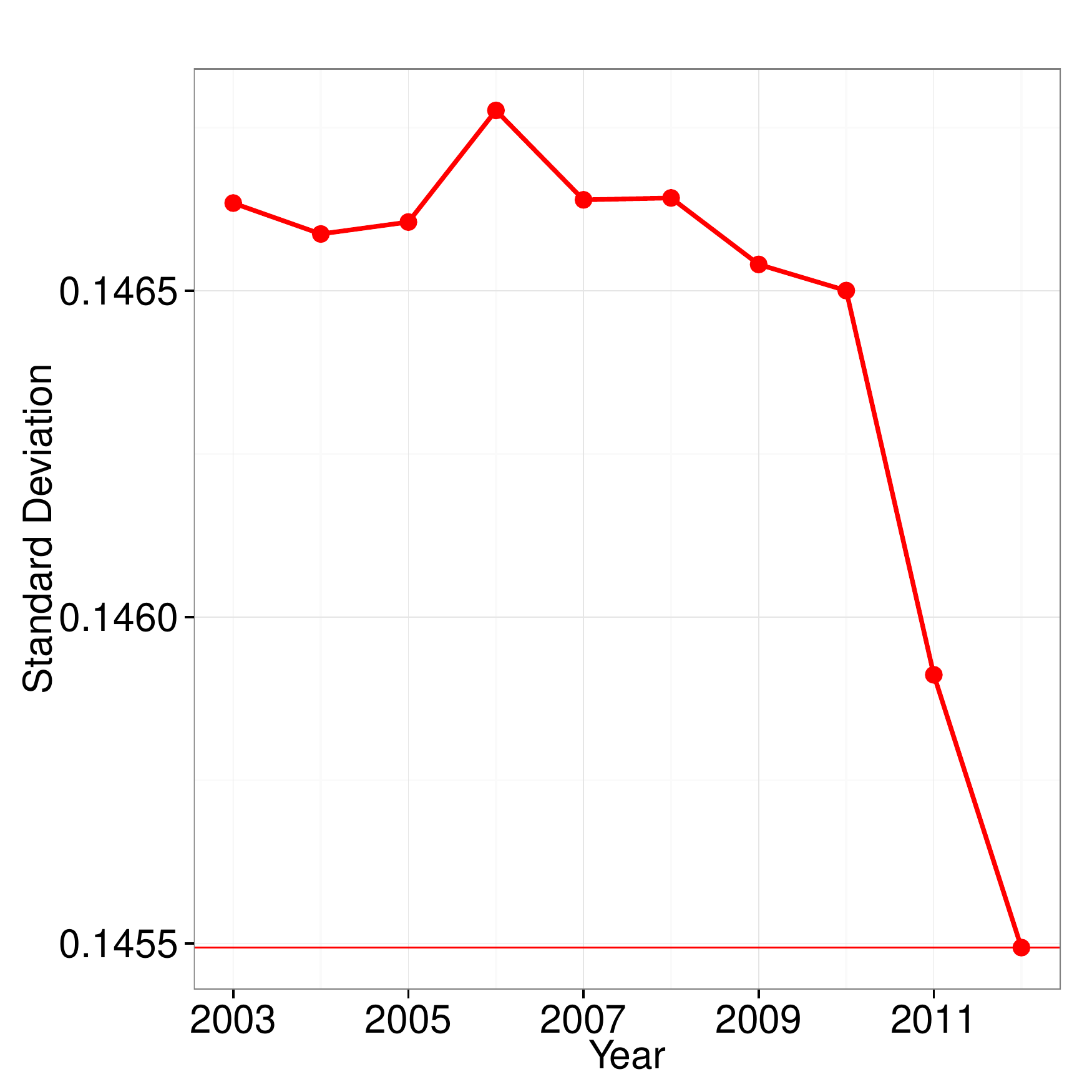}
			\subcaption{2012}
			\label{fig_neg_2012}
		\end{minipage}
		\caption{Standard deviation of each $y_{t'|t}^{neg}$s for years 2003 to 2012.  The horizontal red line denotes the standard deviation in the year analyzed (i.e., $t$).}
		\label{fig4s}
	\end{figure}

\begin{figure}[!h]
		\begin{minipage}[b]{0.243\linewidth}
			\centering
			\includegraphics[keepaspectratio, scale=0.2]
			{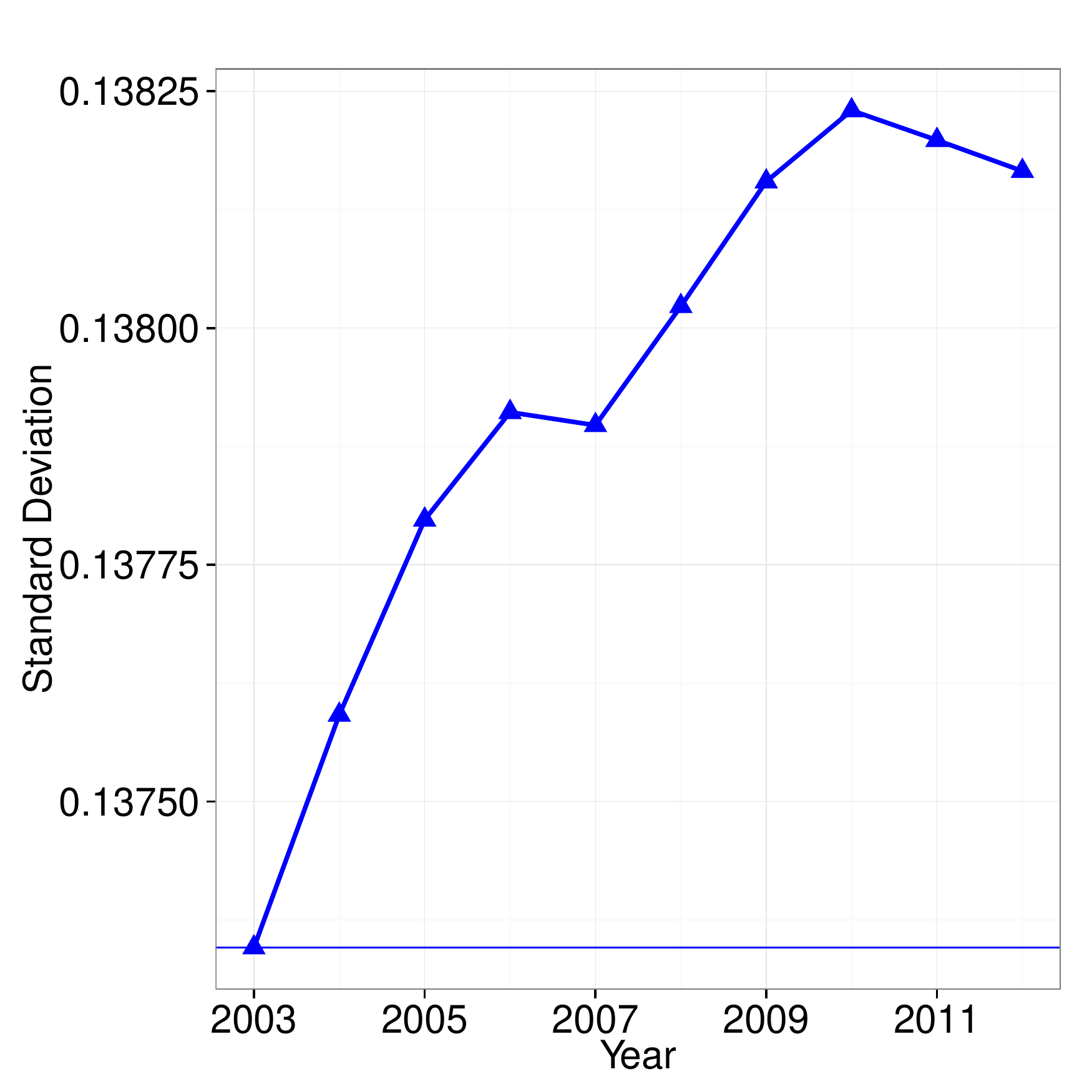}
			\subcaption{2003}
			\label{fig_pos_2003}
		\end{minipage}
		\begin{minipage}[b]{0.243\linewidth}
			\centering
			\includegraphics[keepaspectratio, scale=0.2]
			{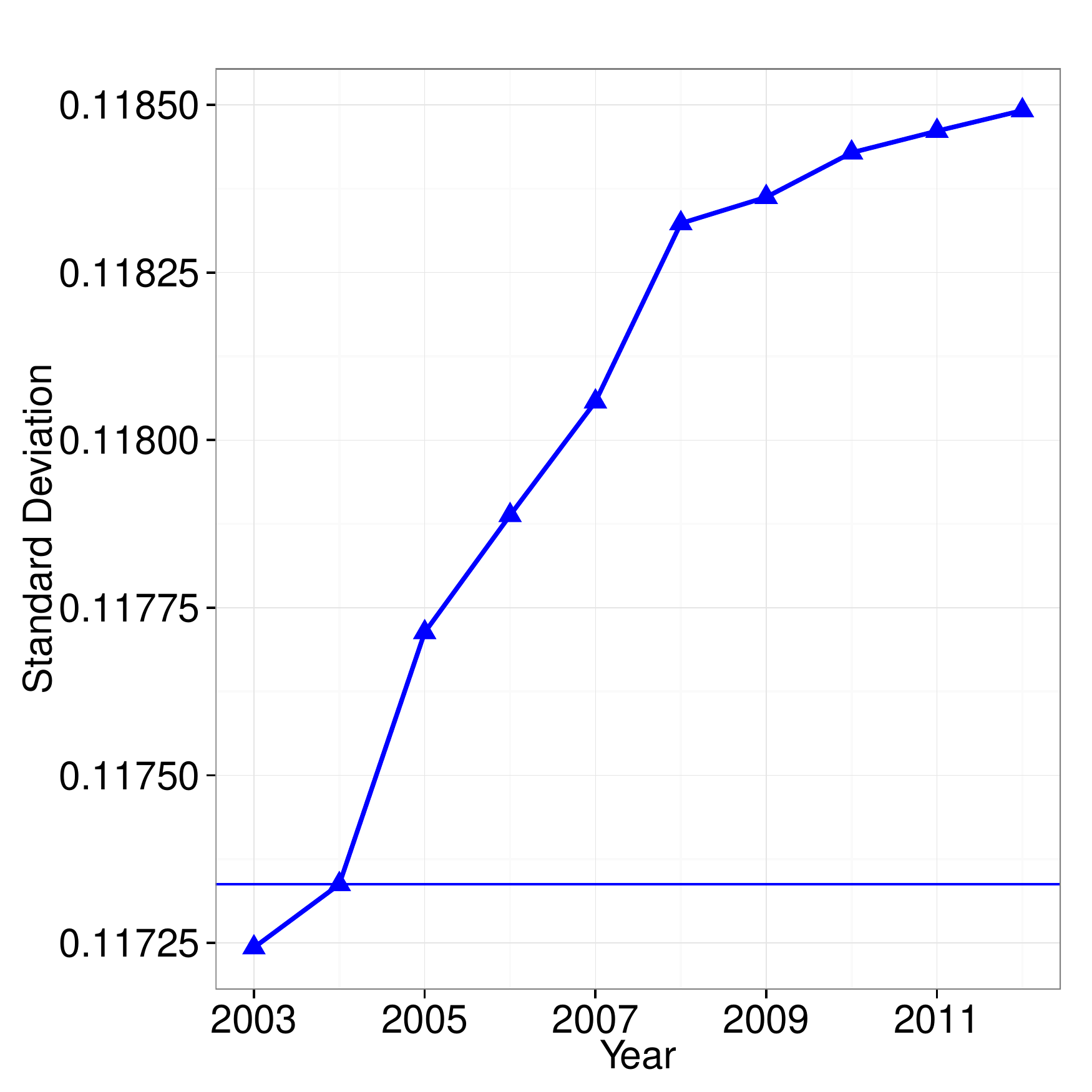}
			\subcaption{2004}
			\label{fig_pos_2004}
		\end{minipage}
		\begin{minipage}[b]{0.243\linewidth}
			\centering
			\includegraphics[keepaspectratio, scale=0.2]
			{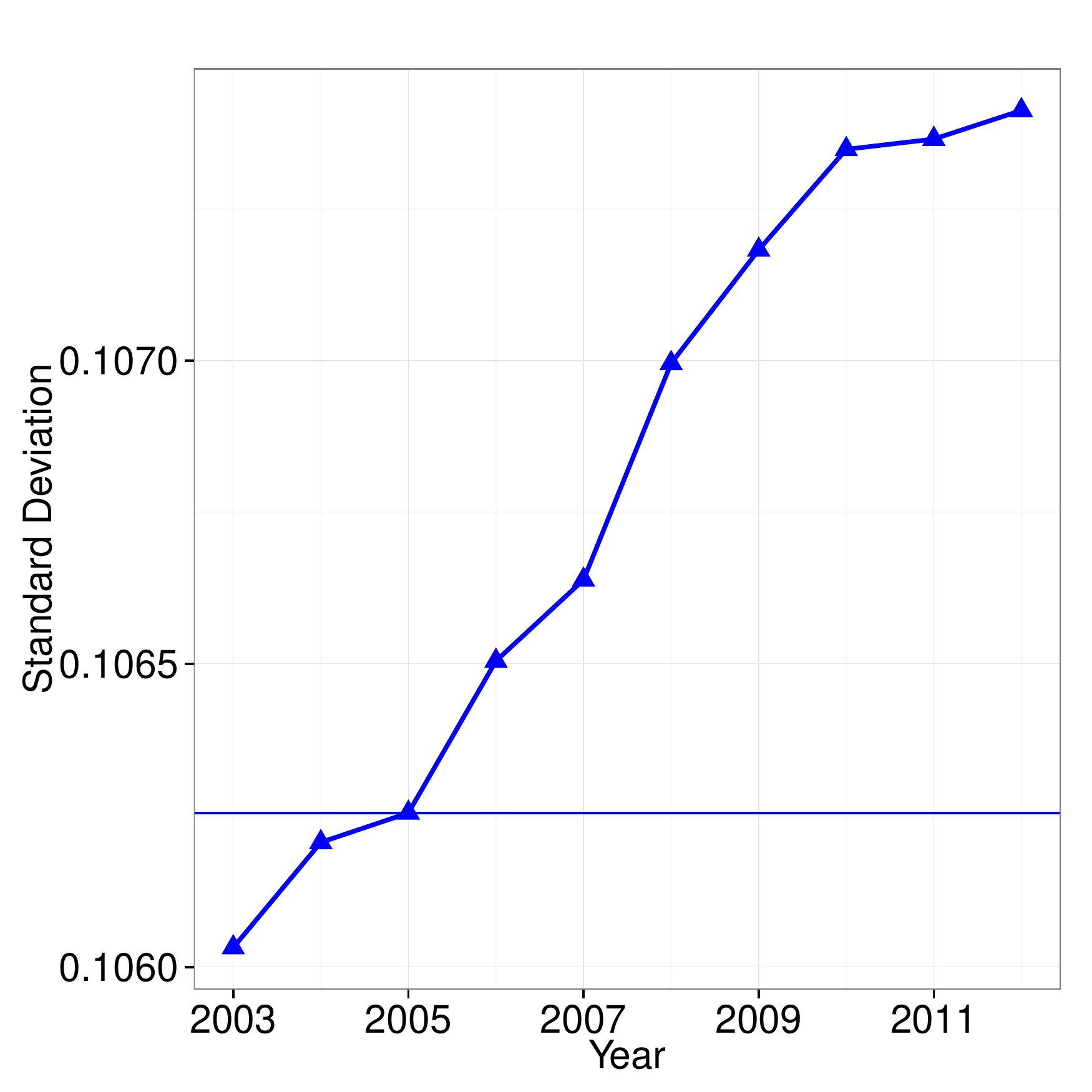}
			\subcaption{2005}
			\label{fig_pos_2005}
		\end{minipage}
		\begin{minipage}[b]{0.243\linewidth}
			\centering
			\includegraphics[keepaspectratio, scale=0.2]
			{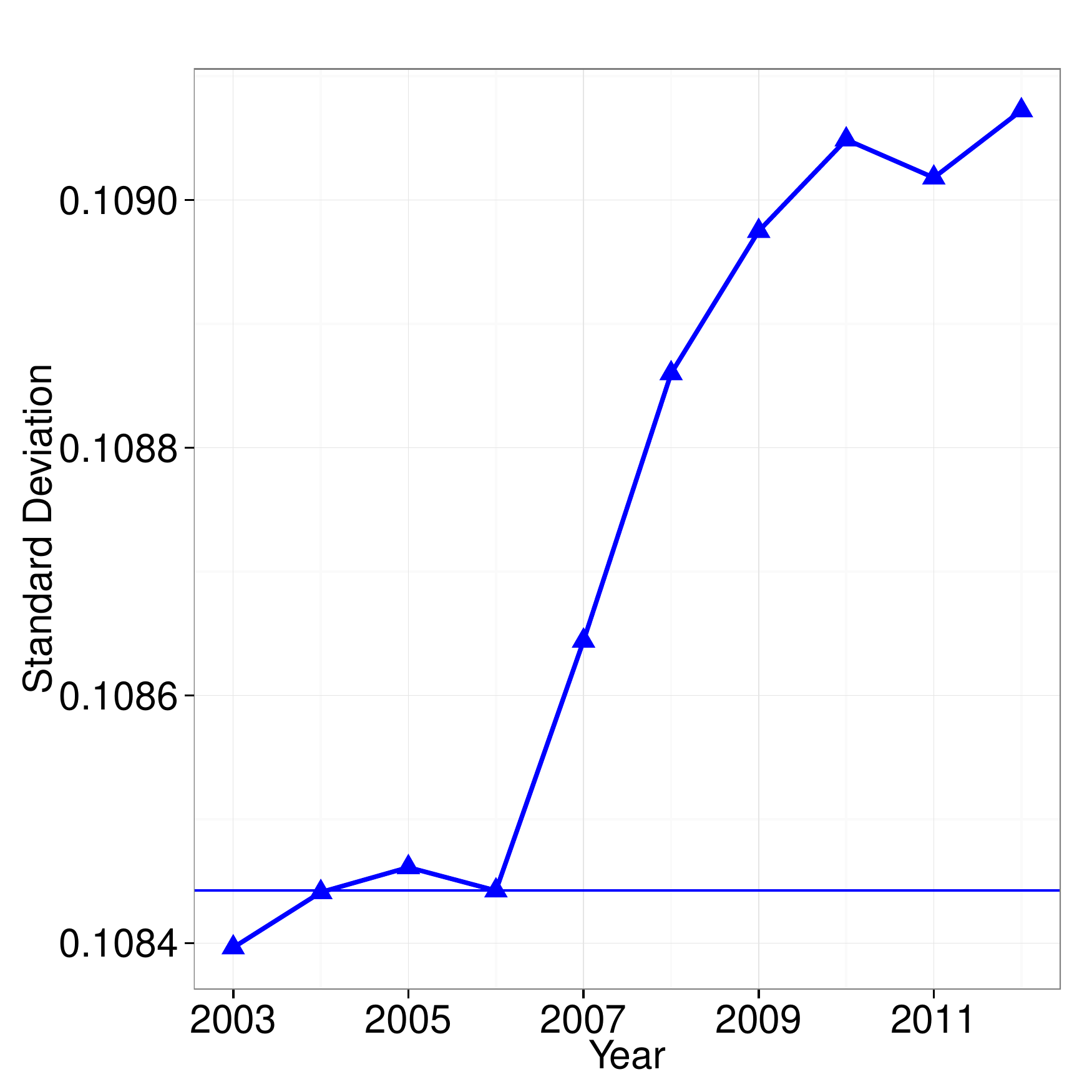}
			\subcaption{2006}
			\label{fig_pos_2006}
		\end{minipage}\\
		\begin{minipage}[b]{0.243\linewidth}
			\centering
			\includegraphics[keepaspectratio, scale=0.2]
			{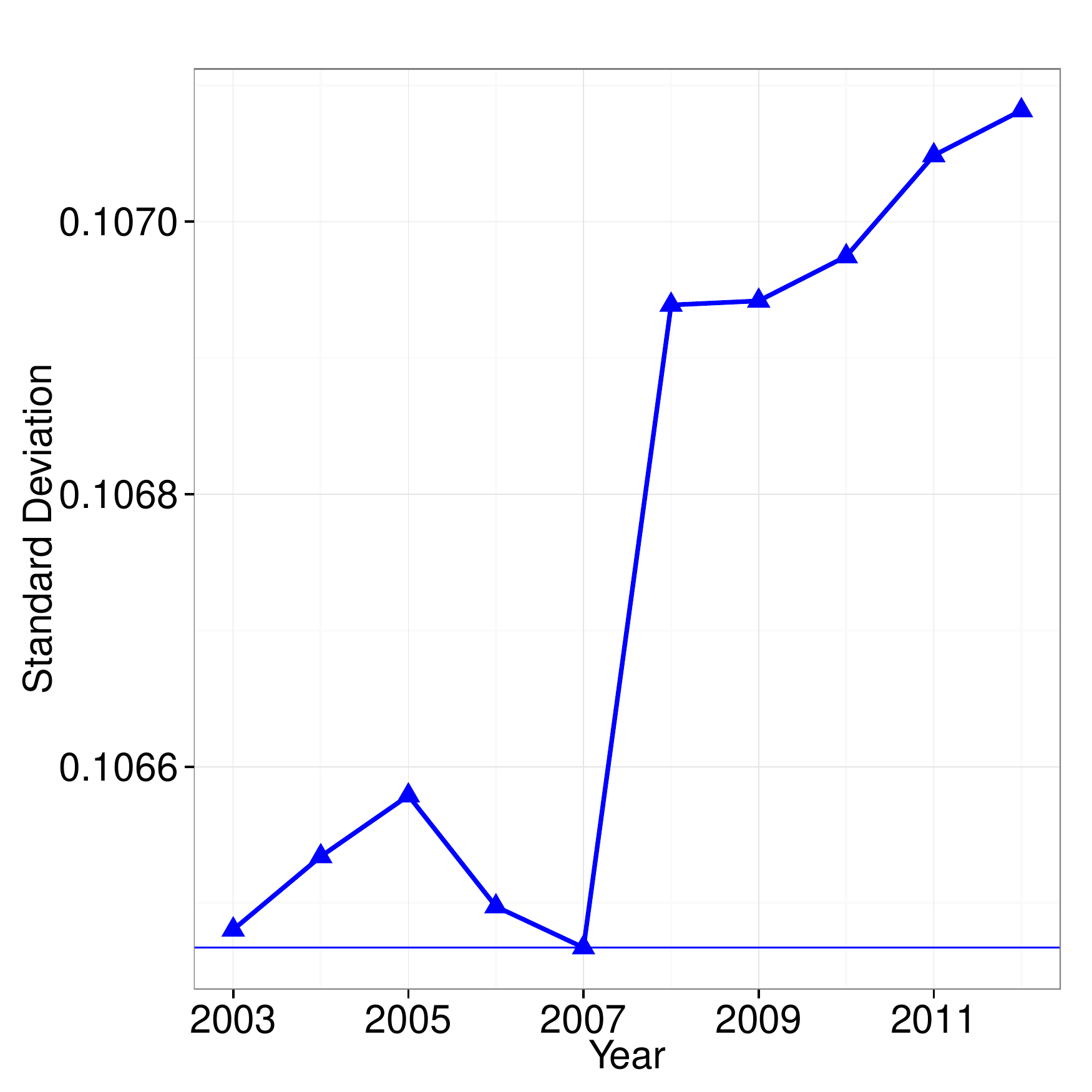}
			\subcaption{2007}
			\label{fig_pos_2007}
		\end{minipage}
		\begin{minipage}[b]{0.243\linewidth}
			\centering
			\includegraphics[keepaspectratio, scale=0.2]
			{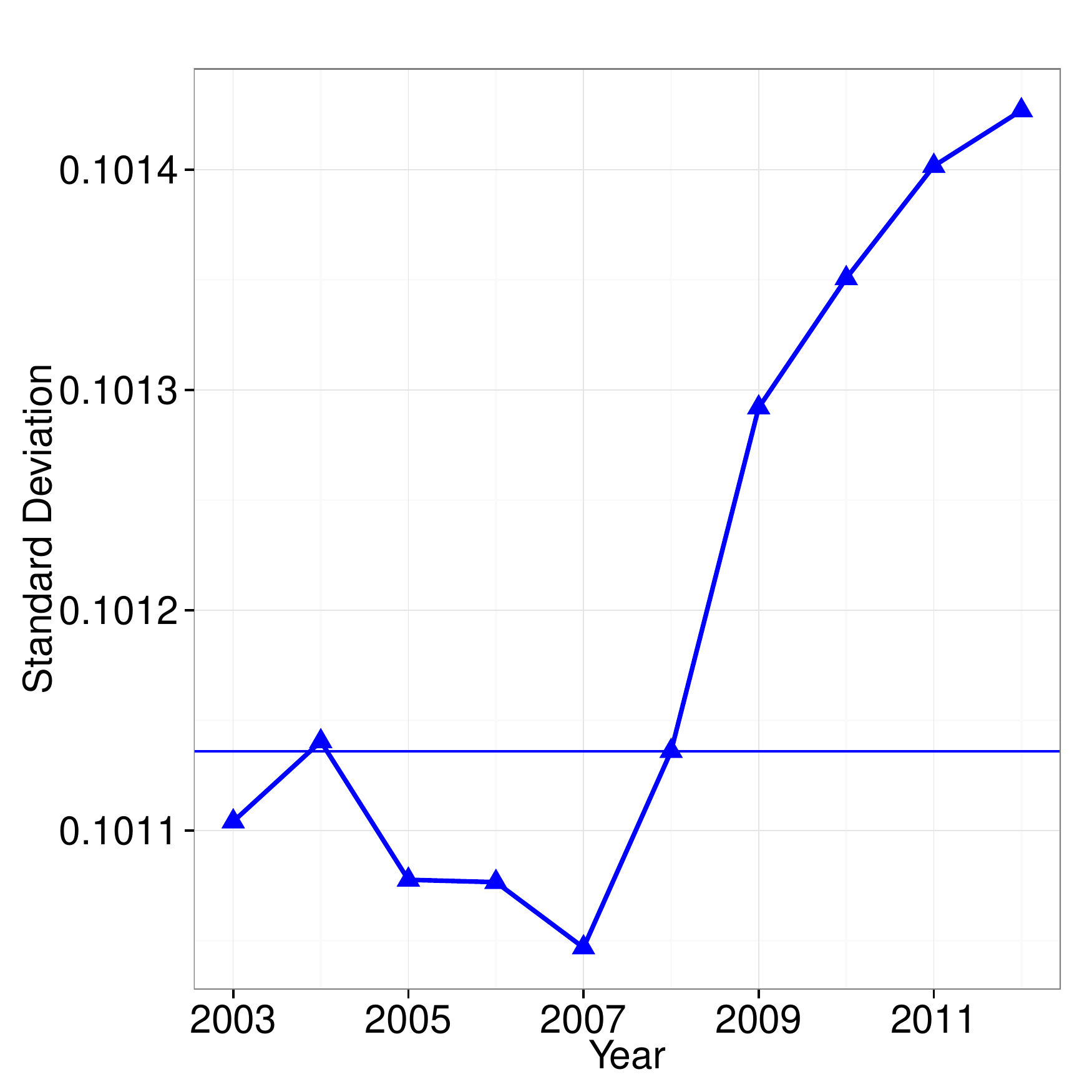}
			\subcaption{2008}
			\label{fig_pos_2008}
		\end{minipage}
		\begin{minipage}[b]{0.243\linewidth}
			\centering
			\includegraphics[keepaspectratio, scale=0.2]
			{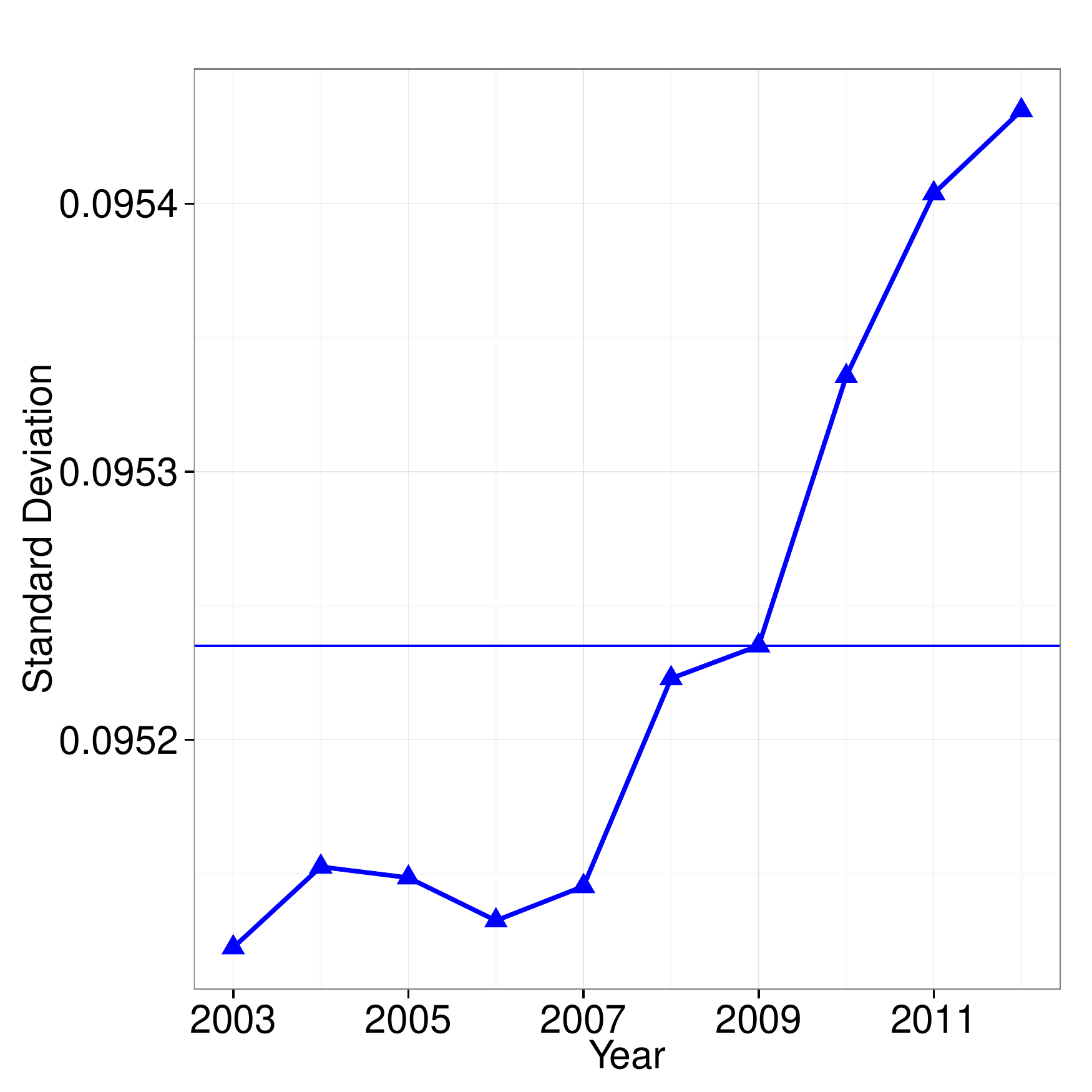}
			\subcaption{2009}
			\label{fig_pos_2009}
		\end{minipage}
		\begin{minipage}[b]{0.243\linewidth}
			\centering
			\includegraphics[keepaspectratio, scale=0.2]
			{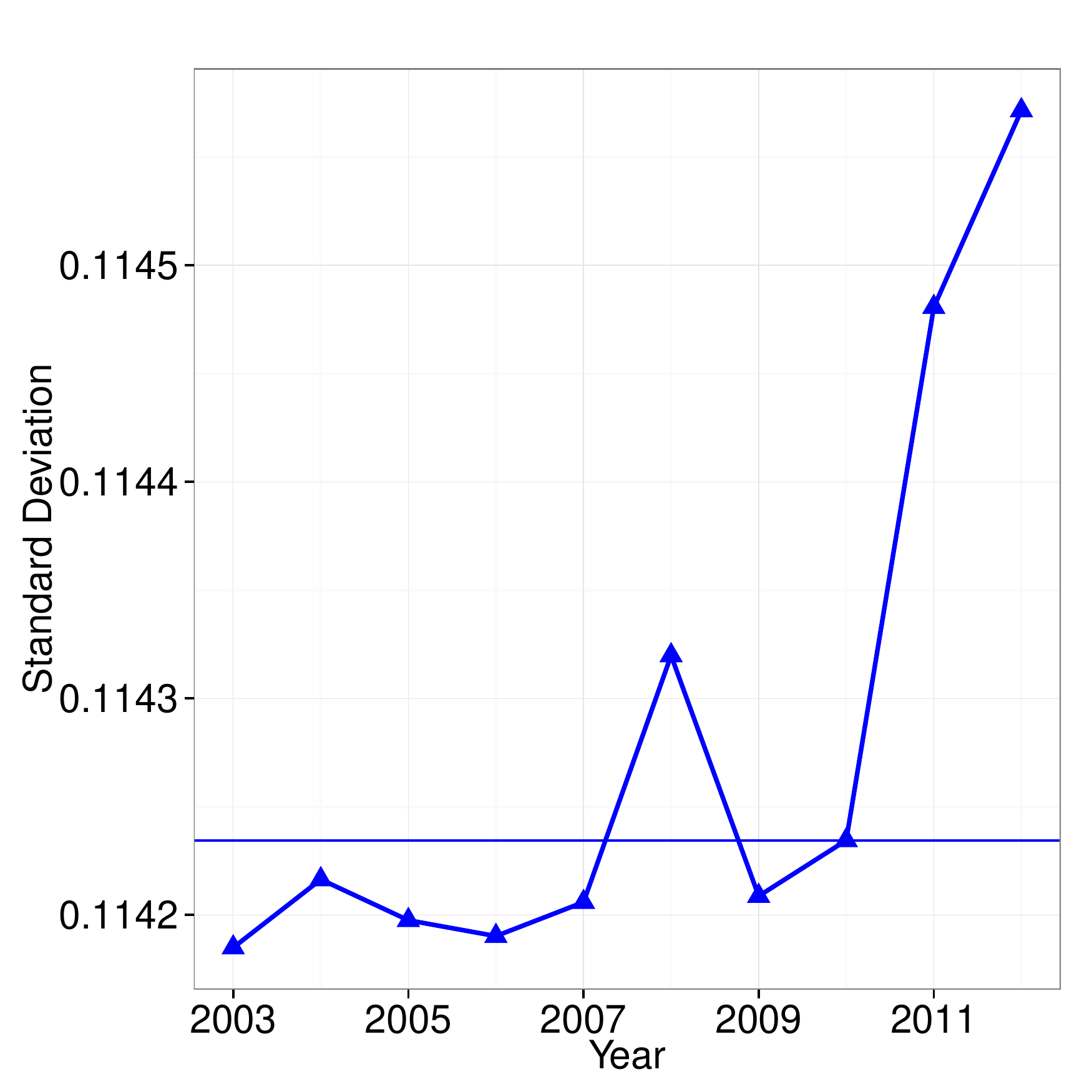}
			\subcaption{2010}
			\label{fig_pos_2010}
		\end{minipage}\\
		\begin{minipage}[b]{0.243\linewidth}
			\centering
			\includegraphics[keepaspectratio, scale=0.2]
			{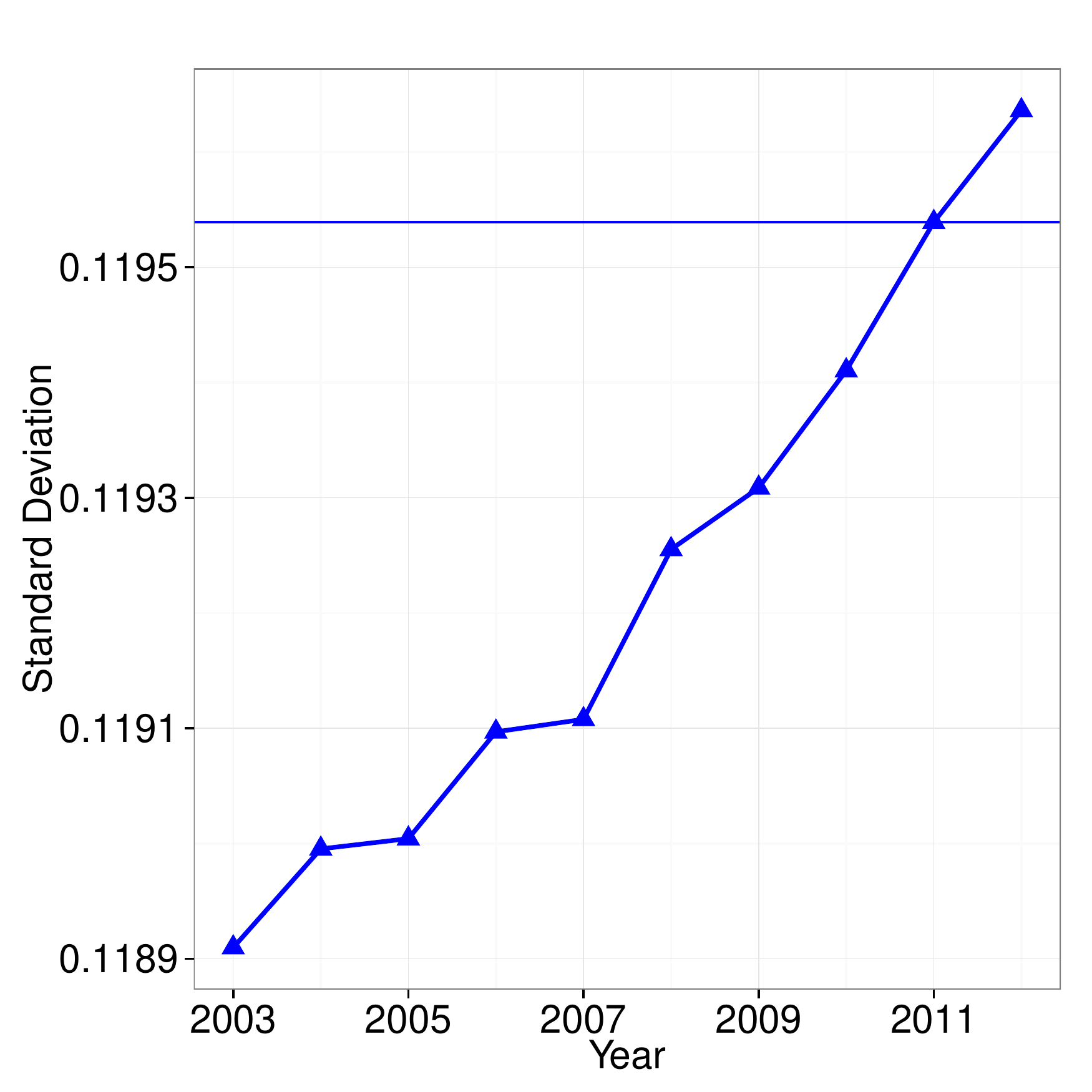}
			\subcaption{2011}
			\label{fig_pos_2011}
		\end{minipage}
		\begin{minipage}[b]{0.243\linewidth}
			\centering
			\includegraphics[keepaspectratio, scale=0.2]
			{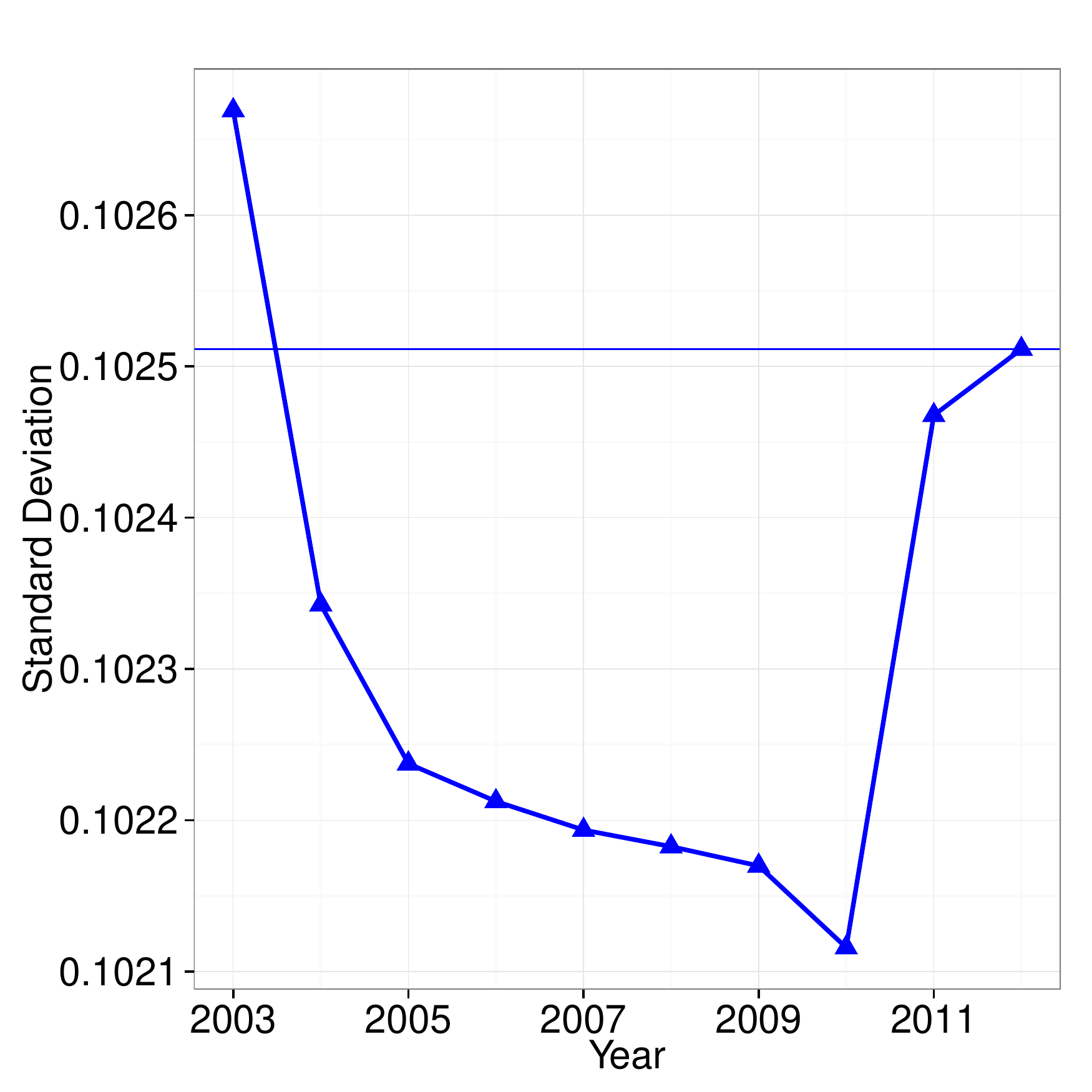}
			\subcaption{2012}
			\label{fig_pos_2012}
		\end{minipage}
		\caption{Standard deviation of each $y_{t'|t}^{pos}$s for years 2003 to 2012.  The horizontal blue line denotes the standard deviation in year $t$.}
		\label{fig5s}
	\end{figure}

	Comparing $y_{t'|t}^{pos}$ and $y_{t'|t}^{neg}$ for different years enables us to compare the propagation (avoidance) performance of each network in the face of positive and negative shocks that arrived in year $t$.  Figures~\ref{fig4s} and ~\ref{fig5s} show the results of comparing the standard deviation of $y_{t'|t}^{pos}$ and $y_{t'|t}^{neg}$ for all years.  It can be seen that the current network is often the best network configuration, which optimizes both the propagation of positive shocks and the avoidance of negative shocks compared with past networks.  Furthermore, we see that for positive shocks, the future network is often better than the current network in the sense that it propagates positive shocks better than the current network.  We also note that the improvement caused by rewiring the network just after the shock has arrived is higher for negative shocks than for positive shocks.

	This is quite an interesting result, and is worth elaborating.  The main reason is the asymmetry between forming and severing links.  Severing a link, and often switching to better (but not necessarily the best) nodes, is easier than forming a link targeting good (if not the best) nodes facing positive shocks.  This is because the latter requires additional search costs and negotiation time for the two firms to reach agreement.  Further, because of the existence of layers (or a hierarchical structure) in the network, creating a path to distant nodes with which one is unable to form a direct link is a complex task that requires decisions by one's neighbors.  For example, if a firm wants to buy automobile parts that use a certain high-quality metal, it has to find an automobile parts manufacturer that uses the metal in their own production or wait until some automobile parts manufacturer starts using the metal in their own production.  Given this basic limitation governing the microeconomic link renewal process of firms, link formation can only evolve gradually in response to newly arrived shocks.  The view of local rewiring of links is also shared with works in social networks such as \cite{Mele2010} and \cite{KrHa14s}.

	If there was a hypothetical social planner that could rewire all the network structures in an economy to an optimal state, the behavior summarized in this section would not take place.  However, in reality, microscopic connectivity patterns are determined by each agent's decisions to avoid negative shocks and share positive shocks.  These decisions are made based on local information which each firms gathers without having access to the full picture of the global state of the network.  Moreover, apart from the fact that they only have access to local information, there is asymmetry in cost between forming and severing links which also contributes to the gradual process of link renewal.  The analysis performed in this section provides some insights into the gradual evolution process, suggesting how the decentralized myopic decisions of individual firms gradually lead to an improvement in the overall state of the network.

	\section{Network Effect on Aggregate Fluctuations} 
	
	Using the parameters reported in the previous section, we estimate the role of networks in aggregate fluctuations by comparing the average log growth rate of firms (i.e., $y_{t}$) and the average shocks for individual firms (i.e., $e_{t}$).  For each year, we calculate $e_{t}$s by

	\begin{eqnarray} 
	e_{t} := ( I-\beta_{G}G_{t}-\beta_{H}H_{t} )y_{t} - ( \beta_{LG}G_{t-1} + \beta_{LH}H_{t-1} )y_{t-1} - \gamma y_{t-1}.
	\end{eqnarray} 
	
	\noindent The average $e_{t}$ is used as the average shock for individual firms.  We also simulate each firms log growth rate assuming that there was no link renewal during the whole period of study.  This is performed by using equation (9), setting $t'$ as 2003.  The average value of $y_{t'|t}$ is used as the average log growth rate in the counterfactual world assuming that no link renewal took place during the whole period of study.\footnote{To be more precise we are assuming that the network stayed as that of year 2003 during the whole period of study 2004-2012.}

\begin{figure}[!h]
		\begin{center} 
			\includegraphics*[width=.85\textwidth]{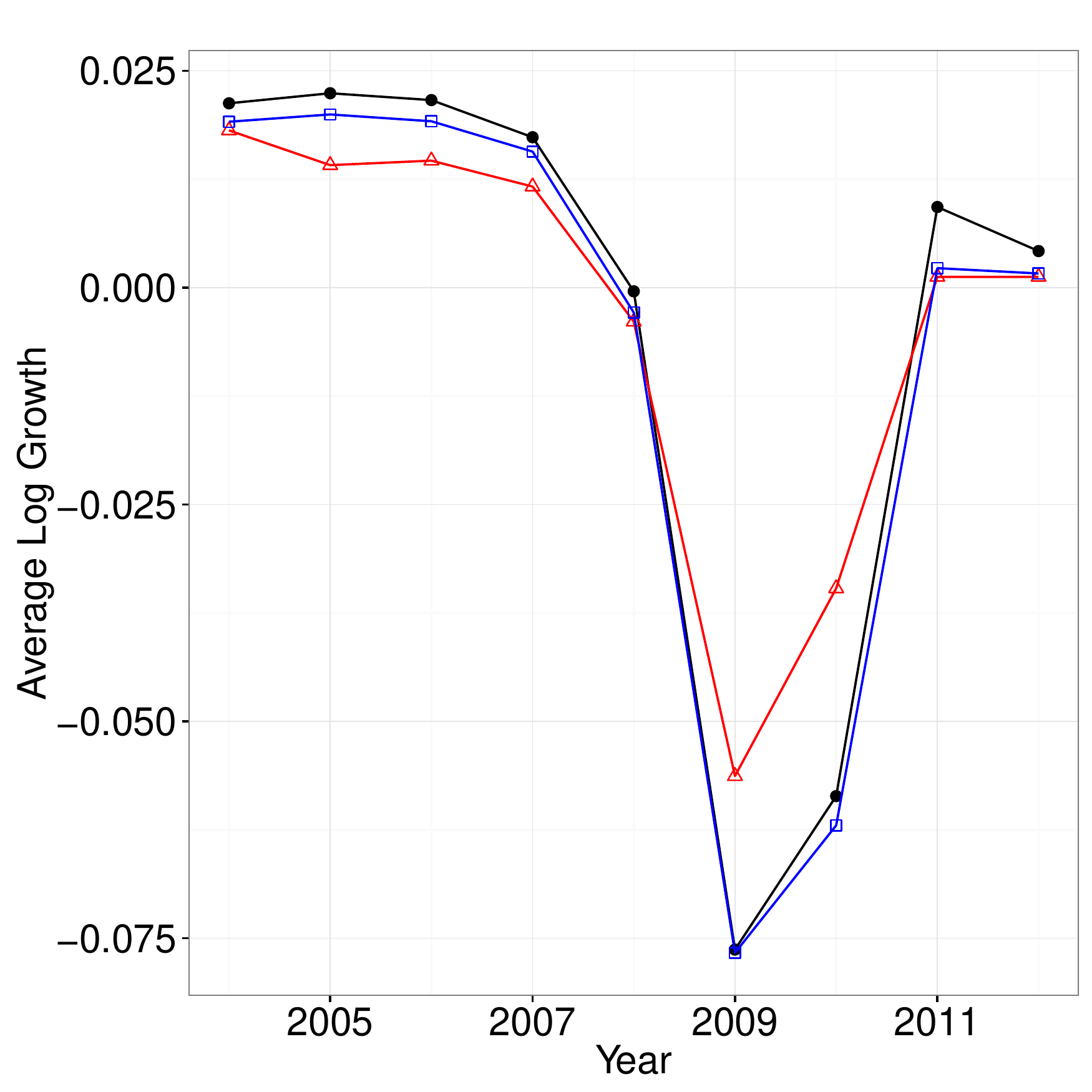} 
			\caption{Time series of average log growth rates (black circles), average shocks for individual firms (red triangles) and simulated average log growth rate assuming that there was no link renewal (blue square) for years 2004 to 2012.} 
			\label{fig:8} 
		\end{center}
	\end{figure}

	Figure~\ref{fig:8} shows the results.  By comparing the case when there is link renewal (black circles) and without link renewal (blue circles), we see that the average log growth rate shifts downwards when there is no link renewal.  This was expected because as was seen in the previous sections link renewal has two effects.  One trying to mitigate negative shocks from propagating and one trying to share positive shocks with their neighboring firms.  In recession period, link renewal is more motivated by the former process making the black circles higher than the blue squares (because by link renewal the network succeeded in mitigating negative shocks).  While in boom period, link renewal is motivated more by the latter process also making the black circle higher than the blue squares (because by link renewal the network succeeded in sharing positive shocks).

\begin{figure}[!h]
		\begin{center} 
			\includegraphics*[width=.85\textwidth]{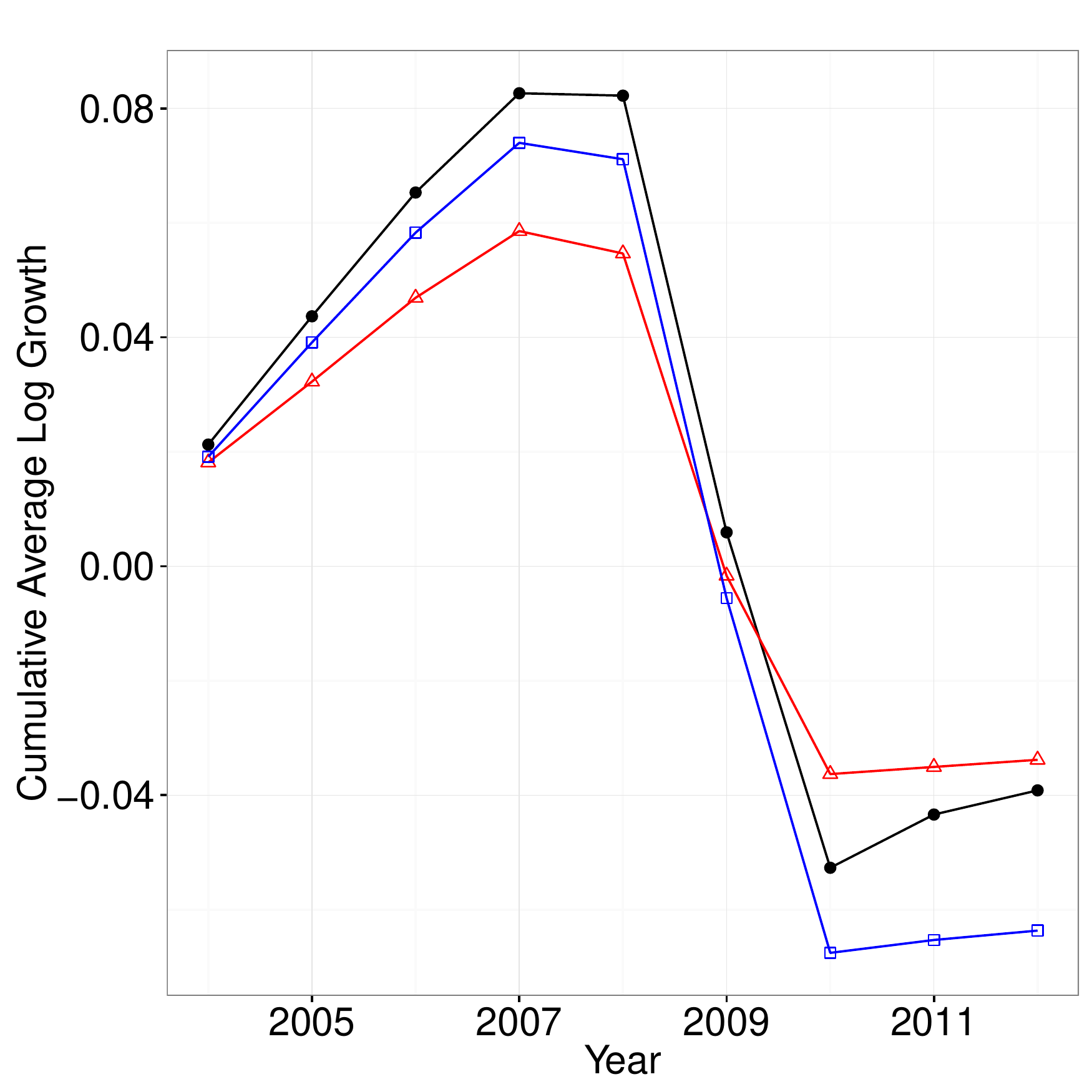} 
			\caption{Time series of cumulative average log growth rate (black circles), cumulative average shocks for individual firms (red triangles) and simulated average log growth rate assuming that there was no link renewal (blue square) for years 2004 to 2012.} 
			\label{fig:9} 
		\end{center}
	\end{figure}

	Figure~\ref{fig:9} shows the cumulative average log growth rate of each of the cases depicted in figure~\ref{fig:8}.  Comparing the cases when link renewal take place (black circles) and when firms are connected and without link renewal (blue square) in figure~\ref{fig:9}, we see that on average firm growth rate is 0.0027 higher when there is link renewal.\footnote{This is calculated by taking the mean of $y_{t} - y_{t'|t}$.}.  Hence we conclude that link renewal has the positive effect of increasing the average log growth rate of an economy by effectively mitigating negative shocks and sharing positive shocks among firms

	We next investigate aggregate fluctuation.  Comparing the two cases when firms are not connected (red triangle) and connected (black circles) in figure~\ref{fig:8}, we see that the average log growth rate tends to fluctuate more when they are connected.  It is worth emphasizing that we only have nine data points in the calculation. Nevertheless, the estimated standard deviation of the fluctuation is 0.023, while that of the original average log growth rate of firms is 0.037.  Thus, the network effect on aggregate fluctuations can be calculated as $1-0.023/0.037$, which is around 37\%.  Note that as discussed in Section 4, the estimated structural parameters provide a lower bound as a result of identification issues concerning measurement errors.  Therefore, we conclude that at least 37\% of the aggregate fluctuations can be explained by the network effect.\footnote{As could be suspected by figure~\ref{fig:8} the number only slightly changes when comparing the case when there is no link renewal to the case when firms are not connected at all}

	It is also worth noting that this figure is similar to that in \cite{Forester2011}, who studied variability in log growth of the IP index in the United States and showed that, after the great moderation, 50\% of the variability in log growth of the IP index could indeed be explained by sectoral linkages.

	\section{Conclusion}

	In order to answer the question concerning the trade-off between propagation of shocks and link renewal in the interfirm buyer--seller network, we provided an empirical analysis on the effect of link renewal on the overall growth rate of an economy.  To this aim we used a firm-level dataset from Japan where we have both network data as well as log growth rate of fimrs over a decade.  Using the unique dataset, we took structural equation modeling to estimate the effect of link renewal.  By means of counterfactual analysis, we first showed that the current network is often the best network configuration which optimizes both the propagation of positive shocks and avoidance of negative shocks compared with previous networks, perhaps reflecting each firms motivation to avoid other's negative shocks and share other's positive shocks.  We then showed that for positive shocks, the future network is often better than the current network in the sense that it propagates positive shocks better than the current network.  This asymmetric behavior was explained by the asymmetry in cost between severing and forming links.  We then provided some evidence that link renewal has a positive effect of increasing the average growth rate of firms at the macroeconomic level answering to the main motivation of the paper.  Last but not least, as a bonus of our structural equation modeling, we also showed that at least 37\% of the aggregate fluctuations can be explained by the network effect.  This is in line with previous research which focused on sectoral linkages such as \cite{Forester2011}.

	\bibliography{Hisano_myref} 
	\bibliographystyle{aea}

\end{document}